%% file: main.tex
\pdfoutput=1
\documentclass[11pt]{article}

\usepackage[final]{acl}
\input{math_commands.tex}

\input{pkgs}
\input{macros}

\usepackage{times}
\usepackage{latexsym}

\usepackage[T1]{fontenc}
\usepackage[utf8]{inputenc}

\usepackage{microtype}

\usepackage{inconsolata}

\usepackage{graphicx}

\title{\technique: Robust Training for Code Generation Models}

\author{
 \textbf{Yuhao Zhang\textsuperscript{1}\thanks{Work done when the author was an intern at Amazon.}},
 \textbf{Shiqi Wang\textsuperscript{1}},
 \textbf{Haifeng Qian\textsuperscript{1}},
 \textbf{Zijian Wang\textsuperscript{1}},
\\
 \textbf{Mingyue Shang\textsuperscript{1}},
 \textbf{Linbo Liu\textsuperscript{1}},
 \textbf{Sanjay Krishna Gouda\textsuperscript{1}},
 \textbf{Baishakhi Ray\textsuperscript{1}},
\\
 \textbf{Murali Krishna Ramanathan\textsuperscript{1}},
 \textbf{Xiaofei Ma\textsuperscript{1}},
 \textbf{Anoop Deoras\textsuperscript{1}}
\\
\\
 \textsuperscript{1}AWS AI Labs,
\\
 \small{
   \textbf{Correspondence:} \href{yhzhangh@amazon.com}{yhzhangh@amazon.com}
 }
}

\begin{document}
\maketitle
\begin{abstract}
% general version
Code generation models are not robust to small perturbations, which often lead to incorrect generations and significantly degrade the performance of these models.
Although improving the robustness of code generation models is crucial to enhancing user experience in real-world applications, existing research efforts do not address this issue.
To fill this gap, we propose \technique, a framework to improve the robustness of code generation models, generalizing a large variety of code perturbations to enrich the training data and enabling various robust training strategies, mixing data augmentation, batch augmentation, adversarial logits pairing, and contrastive learning, all carefully designed to support high-throughput training. 
Extensive evaluations show that we increase the average robust pass rates of baseline CodeGen models from 14.79 to 21.74.
Notably, we decrease the robustness drop rate from 95.02\% to 54.95\% against code-syntax perturbations.
% 1-(2.58/54.91+2.43/46.22+1.46/29.24)/3
% 1-(27.70/53.16+21.92/45.04+10.67/31.04)/3
% \shiqi{should we say the numbers in robust drop \% or relative percentage improvements? The improvement has a maximal limit till nominal.}
% \qianhf{how about: Notably, accuracy drop due to syntax perturbations is improved from \#\#\% to \#\#\%.}
\end{abstract}

\input{sections/introduction}
\input{sections/related_works}
\input{sections/problem_def}
\input{sections/approach}
\input{sections/evaluation}
\section{Conclusion}
We propose \technique, a framework to improve the robustness of code generation models, generalizing a large variety of code perturbations to enrich the training data and enabling various robust training strategies.
Our approach significantly enhances the model robustness and surpasses the sub-optimal results of data augmentation. 
Notably, our approach significantly improves the robustness under code-syntax perturbations, the type of perturbation that hurts the model robustness the most.
Our ablation studies show that ContraSeq, the CL objective used in previous work for MLM, has negligible robustness improvements on CLM.

\section{Limitations and Future Work}
We foresee many future improvements to this paper. 
First, ALPD and ContraName primarily target function and variable rename perturbations but are not general enough to handle arbitrary context-sensitive perturbations.
However, these approaches can be applied to name-entities in general NLP tasks.
Second, the robustness improvement of function-name perturbation on CodeGen-6B and CodeGen-2B is insignificant compared to the baseline, necessitating unique strategies to overcome this limitation.
Thirdly, our evaluation is limited to the CodeGen model architecture and primarily uses popular benchmarks like HumanEval and MBPP. 
However, we have assessed our approach across three different sizes of CodeGen models to illustrate its generalizability.
Furthermore, it is important to note that our perturbed training dataset is generated based on real-world programs from the Stack v1.2 dataset. By training our models on a dataset that follows a real-world program distribution, we hypothesize that models trained using our approach can generalize effectively to other real-world coding benchmarks.

\bibliography{iclr2024_conference}

\newpage
\appendix

\input{sections/appendix}

\end{document}

%% file: math_commands.tex
\usepackage{amsmath,amsfonts,bm}

\def\eqref#1{equation~\ref{#1}}
\def\1{\bm{1}}

\DeclareMathAlphabet{\mathsfit}{\encodingdefault}{\sfdefault}{m}{sl}
\SetMathAlphabet{\mathsfit}{bold}{\encodingdefault}{\sfdefault}{bx}{n}

\newcommand{\E}{\mathbb{E}}

\newcommand{\R}{\mathbb{R}}

\newcommand{\KL}{D_{\mathrm{KL}}}

%% file: pkgs.tex
\usepackage[group-separator={,}]{siunitx}
\usepackage{booktabs} % For formal tables
\usepackage{color}
\usepackage{listings}
\usepackage{enumitem}
\usepackage{threeparttable}
\usepackage{multirow}% http://ctan.org/pkg/multirow
\usepackage{hhline}% http://ctan.org/pkg/hhline
\usepackage{subcaption}
\usepackage{array}
\usepackage{pifont}
\usepackage{booktabs}
\usepackage{dsfont}
\usepackage{microtype}
\usepackage{graphicx}
\usepackage{stmaryrd}
\usepackage{amsmath}

\usepackage{amssymb}
\usepackage{amsthm}
\usepackage{url}
\usepackage{bbm}
\usepackage{xspace}
\usepackage[normalem]{ulem}
\usepackage{booktabs, siunitx}
\usepackage{xcolor}
\usepackage{colortbl}
\usepackage{tkz-euclide}
\usepackage[framemethod=tikz]{mdframed}
\usepackage{empheq}
\usepackage[most]{tcolorbox}
\usepackage{adjustbox}
\usepackage{wrapfig}
\usepackage{pgfplots}
\usepgfplotslibrary{groupplots}
\usepackage{pgfplotstable}

\usepackage{balance}
\expandafter\def\expandafter\UrlBreaks\expandafter{\UrlBreaks%  save the current one
  \do\a\do\b\do\c\do\d\do\e\do\f\do\g\do\h\do\i\do\j%
  \do\k\do\l\do\m\do\n\do\o\do\p\do\q\do\r\do\s\do\t%
  \do\u\do\v\do\w\do\x\do\y\do\z\do\A\do\B\do\C\do\D%
  \do\E\do\F\do\G\do\H\do\I\do\J\do\K\do\L\do\M\do\N%
  \do\O\do\P\do\Q\do\R\do\S\do\T\do\U\do\V\do\W\do\X%
  \do\Y\do\Z}

%% file: macros.tex
\newcommand{\todored}[1]{{\color{red} \{TODO: {#1}\}}}

\newcommand{\code}[1]{{\tt {#1}}}

\expandafter\def\expandafter\UrlBreaks\expandafter{\UrlBreaks%  save the current one
  \do\a\do\b\do\c\do\d\do\e\do\f\do\g\do\h\do\i\do\j%
  \do\k\do\l\do\m\do\n\do\o\do\p\do\q\do\r\do\s\do\t%
  \do\u\do\v\do\w\do\x\do\y\do\z\do\A\do\B\do\C\do\D%
  \do\E\do\F\do\G\do\H\do\I\do\J\do\K\do\L\do\M\do\N%
  \do\O\do\P\do\Q\do\R\do\S\do\T\do\U\do\V\do\W\do\X%
  \do\Y\do\Z}

\definecolor{codegreen}{rgb}{0,0.6,0}
\definecolor{codegray}{rgb}{0.5,0.5,0.5}
\definecolor{codepurple}{rgb}{0.58,0,0.82}
\definecolor{backcolour}{rgb}{1,1,1}
 
\lstdefinestyle{mystyle}{
    backgroundcolor=\color{backcolour},   
    commentstyle=\color{codegreen},
    keywordstyle=\color{magenta},
    numberstyle=\tiny\color{codegray},
    stringstyle=\color{codepurple},
    basicstyle=\ttfamily\footnotesize,
    breakatwhitespace=false,         
    breaklines=true,                 
    captionpos=b,                    
    keepspaces=true,                 
    numbers=left,                    
    numbersep=5pt,                  
    showspaces=false,                
    showstringspaces=false,
    showtabs=false,           
    tabsize=2,
    xleftmargin=1.5em
}

\newtheorem{example}{Example}[section]

\newcommand{\technique}{CodeFort\xspace}
\newcommand{\perturbation}{\pi}
\newcommand{\trans}{T}
\newcommand{\batch}{B}
\newcommand{\pt}[1]{\tilde{#1}}

\newcommand{\bfx}{\mathbf{x}}
\newcommand{\bfm}{\mathbf{m}}
\newcommand{\bfh}{\mathbf{h}}

\newcommand{\bfu}{\mathbf{u}}

\newcommand{\bernoulli}[1]{\mathrm{Bernoulli}(#1)}
\newcommand{\mask}{\bfm}
\newcommand{\drop}[1]{\mathrm{Dp}(#1)}
\newcommand{\unptb}{U}
\newcommand{\freeptb}{F}
\newcommand{\senptb}[1]{S#1}

\newcommand{\trpoint}[2]{#1^{#2}}
\newcommand{\cor}[2]{\mathrm{Cor}(#1+#2)}
\newcommand{\loss}[0]{\mathcal{L}}
\newcommand{\lossclm}[0]{\loss_\mathrm{CLM}}
\newcommand{\lossdataaug}[0]{\loss_\mathrm{DA}}
\newcommand{\lossmdataaug}[0]{\loss_\mathrm{MDA}}
\newcommand{\lossmbatchaug}[0]{\loss_\mathrm{MBA}}
\newcommand{\lossalp}[0]{\loss_\mathrm{ALP}}
\newcommand{\lossalpdrop}[0]{\loss_\mathrm{ALPD}}
\newcommand{\losscontraseq}[0]{\loss_\mathrm{CSeq}}
\newcommand{\losscontratoken}[0]{\loss_\mathrm{CTok}}
\newcommand{\losscontraname}[0]{\loss_\mathrm{CName}}

\newcommand{\npsk}{NP@1\xspace}
\newcommand{\rpsk}{RP$_{10}$@1\xspace}
\newcommand{\rpskf}{RP$_{5}$@1\xspace}
\newcommand{\rperct}{Robust\%\xspace}
\newcommand{\rdrop}{Drop\%\xspace}

\lstdefinelanguage{prompt}{
  basicstyle=\ttfamily\footnotesize,
  moredelim=[is][\textbf]{[[}{]]},
  moredelim=[is][\color{red}]{@@}{@@},
  moredelim=[is][\color{blue}]{~~}{~~},
  breaklines=true,
  postbreak=\mbox{{$\hookrightarrow$}},
  breakindent=0pt,
  frame=lines,
  numbers=none, 
  xleftmargin=0ex,
}

%% file: sections/introduction.tex
\section{Introduction}

% The prominence of code generation models in the contemporary realm of AI applications is undeniable \zijian{nit: reword "undeniable" and try writing it positively. }
Code generation models~\citep{starcoder, codegen2,incoder,wizardcoder,codellama} have demonstrated impressive performance in generating code from natural language descriptions, completing sections of code, and even tackling complex coding contest challenges. 
These models can potentially assist software engineers and increase their productivity. 
% Today, there are numerous commercially available options that leverage AI for the purpose of code generation, underscoring its significance in the modern software development landscape.

\begin{figure}
    \centering
    \includegraphics[width=0.83\columnwidth]{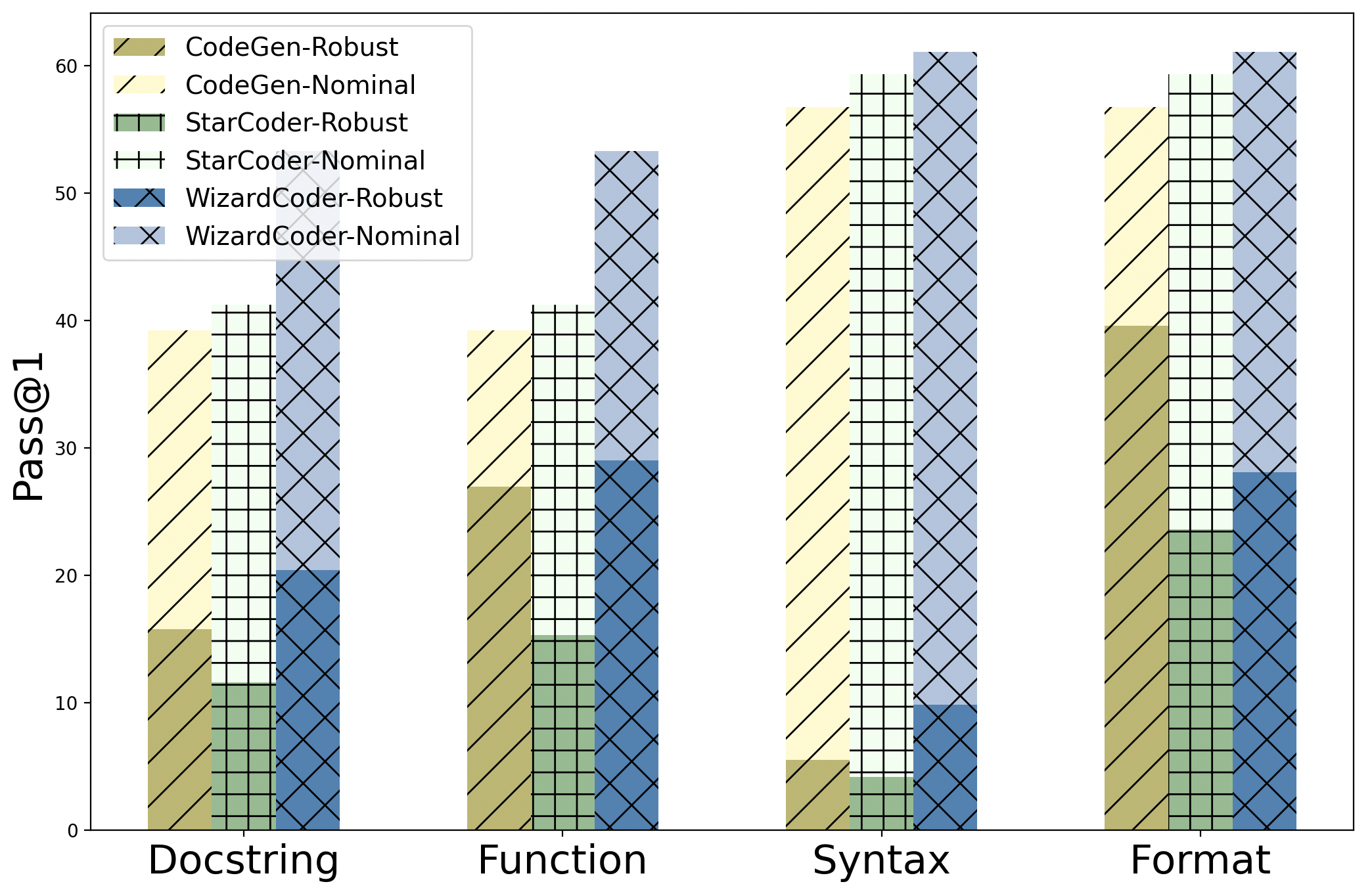}
    \vspace{-0.5em}
    \caption{Performance drop of the state-of-the-art public code models on four classes of code perturbations. (See Table~\ref{tab:motivation} in the Appendix for details.)}
    \label{fig:motivation}
    \vspace{-1.5em}
\end{figure}

However, code generation models are not robust to minor perturbations in the input prompts (e.g., inserting whitespaces/typos in docstrings or substituting variable names in code), i.e., they often generate incorrect outputs, thus significantly degrading their impressive performance on nominal prompts and hurting user experience when deployed in real-world applications~\citep{recode}. 
Figure~\ref{fig:motivation} shows that the performance of the state-of-the-art public code models~\citep{codegen,starcoder,wizardcoder} significantly declines under semantic-preserving program transformations, particularly under code-syntax perturbations.
Thus, it is crucial to improve the robustness of models before they are universally deployed.
% In this paper, we focus on answering the following question,
% \begin{center}
%     \emph{Can we design a framework to train robust code generation models?}
% \end{center}

% \input{tables/motivation_small}.

Despite extensive research efforts to improve the robustness of code-related tasks, beyond code generation, %other than code generation 
such as vulnerability prediction, clone detection, and code summarization, existing work has not tackled two unique challenges of improving the robustness of code generation models, primarily trained using casual language modeling (CLM).

% Define a counter for the challenges
\newcounter{challengecounter}
\setcounter{challengecounter}{0} % Initialize the counter

% Define the new environment with an argument
\newenvironment{challenge}[1]{%
    \refstepcounter{challengecounter}% Increment the counter and make it referencable
    \paragraph{Challenge \arabic{challengecounter}: #1}%
}

% challenge of existing approaches
\begin{challenge}{Distinct Robustness Definition}
\label{challenge1}
    Unlike traditional classification tasks like vulnerability detection, where models produce a single classification, code generation models generate sequences, leading to a shift in the definition of robustness for certain perturbations. 
In code generation, model robustness is defined by generating a \emph{coherent} output given an input perturbation. In contrast, in classification tasks, models are expected to maintain the \emph{same} classification before and after perturbation. 
For instance, if a variable \code{i} is renamed to \code{b} in the input prompt (as illustrated in Figure~\ref{fig: ptb_syntax_format_exp}), a robust code generation model should generate completions with the variable \code{b} instead of the original variable \code{i}. 
This shift necessitates a new category of perturbations for code generation models and corresponding robust training approaches to tackle this new category.
\end{challenge}

\begin{challenge}{Designing Robust Training Approaches} 
\label{challenge2}
As code perturbations can insert dead code or typos into training data, directly training using data augmentation could adversely affect the model performance, leading to issues like generating dead code or typos.
Furthermore, applying more deliberate robust training approaches such as adversarial logits pairing (ALP)~\citep{ALP} and contrastive learning (CL) to CLM presents unique challenges. 
ALP, designed for single-class classification, requires careful alignment between original and perturbed sequences, complicated by potential differences in sequence lengths. 
Although CL has demonstrated efficacy in improving the robustness of code representations in masked language modeling (MLM)~\citep{bertmlm}, its applicability to improving the robustness of code generation models remains unexplored. 
\end{challenge}

\smallskip

To tackle the above two challenges and improve the robustness of code generation models, we introduce a structured definition of code perturbations (Section~\ref{sec: perturb_def}) and design a novel robust training framework named \technique (Section~\ref{sec: framework}). 
% \technique generalizes various code perturbations to enrich the training data and train code generation models with different robust training approaches.

To address Challenge~\ref{challenge1}, we classify existing code perturbations into two categories: \emph{context-free} and \emph{context-sensitive}, based on the formal definition of code perturbations provided in Section~\ref{sec: perturb_def}.
Context-free perturbations, such as the docstring perturbation in Figure~\ref{fig: ptb_docstring_funcname_exp}, follow the traditional notions of robustness, whereas context-sensitive perturbations, such as the code-syntax perturbation in Figure~\ref{fig: ptb_syntax_format_exp}, specific the distinct robustness definition highlighted in Challenge~\ref{challenge1}.
The distinction of these two categories allows \technique to employ different robust training methods according to each category.
% As a result, we propose two novel approaches tailored to context-sensitive robustness: \emph{ALP with name-Dropout (ALPD)} and \emph{name-level CL (ContraName)}.

% easily extend to new perturbations
% which serve as a high-level interface for perturbed dataset generation and down-streaming robust training. 

To address Challenge~\ref{challenge2}, \technique employs example-level and sequence-level pairing to enrich the training set.
These two pairing levels allow 1) a masking mechanism to mask unnatural perturbed tokens and 2) a careful alignment between the original and perturbed token sequences, addressing the crucial challenge of applying ALP and CL to CLM.
% Additionally, we propose a novel \emph{token-level} CL (\emph{ContraToken}) inspired by \citet{contraclm}.
% ContraName and ContraToken empower CLM to discriminate finer-grained representations.

% However, the module does not contain adversarial training approaches because they require to collect gradients and solving discrete optimization problems. Both requirements are time-consuming and cannot be applied to large language models, which usually needs thousands of GPU-hours to train.

We utilize \technique to extensively evaluate various strategies, mixing data augmentation, batch augmentation, ALP, and CL.
Our approach, combining batch augmentation with the masking mechanism, ALP, and ALPD, significantly enhances the model robustness and surpasses the sub-optimal results of data augmentation. 
Notably, our approach significantly improves the robustness under code-syntax perturbations, the type of perturbation that hurts the model robustness the most, as shown in Figure~\ref{fig:motivation}.
Our ablation studies show that ContraSeq, the CL objective used in previous work for MLM, has negligible robustness improvements on CLM.

We summarize our contributions: 
1) a framework, \technique, for improving the robustness of code generation models trained by CLM,
2) designs of robust training approaches, data augmentation, batch augmentation, ALP, and CL tailored to CLM,
3) an extensive evaluation of different robust training approaches,
4) a perturbed training set for future studies on the robustness of code generation models,
and 
5) a surprising finding that the ContraSeq CL
objective, which is known to be beneficial for improving robustness of other code related tasks, has negligible robustness improvements on CLM.
% \todored{add contribution as the first data augmentation.}

%% file: sections/related_works.tex
\section{Related Work}

\input{imgs/examples}

\paragraph{Adversarial Attacks on Code-Related Tasks}
Numerous adversarial attacks~\citep{code_adv1, code_adv2, code_adv3, code_adv5, code_adv6, code_adv7, robust_discrete} have targeted encoder-decoder models in code-related tasks, including classification (e.g., vulnerability prediction) and generation (e.g., code summarization). 
Key methods include CODA~\citep{attack_coda}, which exploits syntactic differences for adversarial example generation; CARROT~\citep{carrot}, employing a lightweight hill climbing for optimization in attacks; and ALERT~\citep{alert}, which creates naturalness-aware attacks using pre-trained models. 
Unlike these approaches, we focus on improving the robustness of \emph{code generation} models trained using \emph{casual language modeling} (CLM). 
We assess our approaches' effectiveness in code generation through ReCode~\citep{recode}, a benchmark for evaluating robustness via semantic-preserving program transformations.

% Many adversarial attacks~\citep{code_adv1, code_adv2, code_adv3, code_adv5, code_adv6, code_adv7, robust_discrete} have been proposed to attack encoder-decoder models on code-related tasks, such as classification tasks like vulnerability prediction 
% %~\citep{vulnerability_pred} 
% and clone detection%~\citep{clone_detection}
% , and generation tasks like code summarization
% %~\citep{code_sum} 
% and comment generation. %~\citep{comment_gen}. 
% Among them, CODA~\citep{attack_coda}, uses the syntactic differences between the target input and reference inputs to guide the generation of adversarial examples.
% CARROT~\citep{carrot} uses a light-weighted hill climbing approach to solve the optimization problem in adversarial attack.
% ALERT~\citep{alert} generates naturalness-aware attack by leveraging the pre-trained models that victim models are fine-tuned on.
% Different from the scope of these adversarial attacks, we focus on improving the robustness of \emph{code generation} models trained using \emph{casual language modeling}. To evaluate the code generation performance of our approach, we use ReCode~\citep{recode} which is a comprehensive robustness evaluation benchmark that contains several semantic-preserving program transformations.
% for code generation models. It contains carefully-designed semantic-preserving program transformations on docstrings, function  names, code syntax, and code format.
% We evaluate our approach on the ReCode benchmark. \myshang{I didn't get the connection between the paragraph and the previous paragraph. }

\paragraph{Robust Training on Code-Related Tasks}
Existing work typically enhances model robustness through data augmentation and adversarial training~\citep{pgd}. 
\citet{robust_ast_hole} refine model representations by feeding only pertinent program parts to the model;
\citet{robust_delta_debugging} use curriculum learning and data augmentation with simplified programs.
They all tend to improve robustness in classification tasks. Unlike these, our focus is on code generation robustness.

% Most of adversarial attacks can be used to improve the robustness of the model by data augmentation and adversarial training~\citep{pgd}. 
% \citet{robust_ast_hole} propose to refine the adversarially trained model representation by feeding only relevant program parts to the model. 
% \citet{robust_delta_debugging} improve the model robustness by curriculum learning and data augmentation with simplified programs~\citep{delta_debugging}.
% These approaches aim to improve robustness on classification tasks, while our approach aims to improve robustness on code generation. 

While \citet{code_adv4_mask} propose random input token masking to lessen dependence on non-robust features, our method selectively masks perturbed tokens during loss calculation to avoid the model generating unnatural perturbations. 

In contrast to ContraCode~\citep{contracode} and ContraBERT~\citep{contrabert}, which apply contrastive learning (CL) to classification and code translation tasks by improving robustness in masked language modeling, 
we focus on the efficacy of CL in decoder-only code generation models. 
Furthermore, directly applying the CL objective of ContraBERT and ContraCode on sequence representations may not cater to CLM, which involves discriminating representations at a finer level than sequence representations.
Notably, this adoption shows negligible robustness improvement on CLM code models (Section~\ref{sec: ablations}). 
Thus, we need to design novel CL objectives tailored to finer granularities.

Although ContraCLM~\citep{contraclm} enhances the discrimination of CLM's representations, it does not target robustness improvement.

%% file: imgs/examples.tex
\begin{figure*}
\centering
\begin{subfigure}[b]{0.47\textwidth}
\centering
\begin{lstlisting}[language=prompt, basicstyle=\ttfamily\tiny]
def largest_divisor(n: int) -> int:
    """ For a given number n, find the largest number that divides n evenly, smaller than n
    >>> largest_divisor(15)
    5
    """
[[===]]
    for i in reversed(range(n)):
        if n % i == 0:
            return i
\end{lstlisting}
\vspace{-1em}
\caption{An original problem in HumanEval. \textbf{\tiny===} separates the prompt and the ground-truth completion.}
\label{fig: ori_exp}
\end{subfigure}\hspace{2em}
\begin{subfigure}[b]{0.47\textwidth}
\centering
\begin{lstlisting}[language=prompt, basicstyle=\ttfamily\tiny]
def @@largestDivisor@@(n: int) -> int:
    """ For a given number n, find the largest number that ~~separate~~ n evenly, ~~modest~~ than n
    >>> @@largestDivisor@@(15)
    5
    """
[[===]]
    for i in reversed(range(n)):
        if n % i == 0:
            return i
\end{lstlisting}
\vspace{-1em}
\caption{A perturbed version of Figure~\ref{fig: ori_exp} by a {\color{red} function-name} and {\color{blue} docstring} perturbation.}
\label{fig: ptb_docstring_funcname_exp}
\end{subfigure}
\begin{subfigure}[b]{0.47\textwidth}
\centering
\begin{lstlisting}[language=prompt, basicstyle=\ttfamily\tiny]
def largest_divisor(n: int) -> int:
    """ For a given number n, find the largest number that divides n evenly, smaller than n
    >>> largest_divisor(15)
    5
    """
    for i in reversed(range(n)):
[[===]]
        if n % i == 0:
            return i
\end{lstlisting}
\vspace{-1em}
\caption{A HumanEval problem includes the first half of the original completion. }
\label{fig: ori_partial_exp}
\end{subfigure}\hspace{2em}
\begin{subfigure}[b]{0.47\textwidth}
\centering
\begin{lstlisting}[language=prompt, basicstyle=\ttfamily\tiny]
def largest_divisor(n: int) -> int:
    """ For a given number n, find the largest number that divides n evenly, smaller than n
    >>> largest_divisor(15)
    5
    """
    for @@b@@ ~~\~~
  in reversed(range(n)):
[[===]]
        if n % @@b@@ == 0:
            return @@b@@
\end{lstlisting}
\vspace{-1em}
\caption{A perturbed version of Figure~\ref{fig: ori_partial_exp} by a {\color{red} code-syntax} and {\color{blue} code-format} perturbation.}
\label{fig: ptb_syntax_format_exp}
\end{subfigure}
\vspace{-0.5em}
    \caption{HumanEval problems under different code perturbations. To achieve a more compact illustration, we merge two code perturbations in one example.}
    \label{fig: examples}
    \vspace{-1em}
\end{figure*}

%% file: sections/problem_def.tex
\section{Problem Definition}
\label{sec: perturb_def}

We address the robustness challenge in a code generation model $f$ trained using Causal Language Modeling (CLM).
% CLM formulates the task as estimating the distribution of sequences. 
% To render this estimation tractable, a common practice of CLM is to approximate the joint distribution of a sequence by multiplying conditional next-token prediction probabilities. \linbo{can be removed to save space}
% In other words, 
CLM predicts the next token in a sequence, and the model can only attend to tokens on the left.
Formally, given a sequence of tokens $\bfx = x_1, \ldots, x_{n}$, the generation model $f$ captures $p_f(\cdot \mid \bfx_{:i})$, representing the conditional probabilities of the $i$-th token given the preceding tokens $\bfx_{:i} = x_1, \ldots, x_{i-1}$. The model is trained on a dataset $D=\{\trpoint{\bfx}{j}\}_{j=1}^m$ using cross-entropy loss $\lossclm(\bfx) = -\sum_{i=1}^{n} \log p_f(x_i \mid \bfx_{:i})$.
% denoting the loss for a single training example $\bfx$.
% \begin{align}
%     \lossclm(\bfx) = -\sum_{i=1}^{n} \log p_f(x_i \mid \bfx_{:i})\label{eq: lossclm}
% \end{align}
% The generation model $f$ determines the most likely next token $\hat{x}_i$ by selecting the token with the highest probability in $p_f(\cdot \mid \bfx_{:i})$ from the entire vocabulary $\alphabet$,
% $\hat{x_i} = \mathrm{argmax}_{v\in \alphabet} p_f(v \mid \bfx_{:i})$.

Utilizing a given decoding strategy, such as greedy or temperature sampling~\citep{nucleus}, the generation model $f$ produces a sequence of tokens by iteratively predicting the next tokens until a specified stop criterion is reached.
We denote $f(\bfx_{:i}) = \hat{\bfx}_{i:}$ as the generated token sequence by $f$.
% as the generation model taking an input $\bfx_{:i}$ and generating an output $\hat{\bfx}_{i:}=\hat{x}_{i}, \ldots, \hat{x}_{\hat{n}}$. 
The terms \emph{prompt}, \emph{completion}, and \emph{ground truth} refer to the input $\bfx_{:i}$, the output $\hat{\bfx}_{i:}$, and the original completion $\bfx_{i:}=x_i, \ldots, x_{n}$, respectively.

In code generation, the token sequence $\bfx$ represents a code snippet. 
% Performance evaluation for code generation models involves execution-based benchmarks like HumanEval~\citep{humaneval} and MBPP~\citep{mbpp}. 
Figure~\ref{fig: ori_exp} shows a problem in HumanEval~\citep{humaneval}.
Each prompt $\bfx_{:i}$ in a problem contains a function signature and a corresponding docstring description, and each ground-truth completion $\bfx_{i:}$ is the correct function implementation. 
Given a completion $\hat{\bfx}_{i:}$ generated by a model $f$,
if the completed function $\bfx_{:i} + \hat{\bfx}_{i:}$ passes all the hidden test cases, the completion is deemed correct, denoted as $\cor{\bfx_{:i}}{\hat{\bfx}_{i:}} = \emph{true}$. Otherwise, if any of the tests fail, $\cor{\bfx_{:i}}{\hat{\bfx}_{i:}} = \emph{false}$.

\subsection{Code Perturbations}
ReCode~\citep{recode} is a comprehensive robustness evaluation benchmark for code generation models containing semantic-preserving code perturbations across four classes: docstring, function-name, code-syntax, and code-format perturbations.
We present examples from these four classes and we encourage readers to refer to the original paper for more detailed descriptions. 
Table~\ref{tab:overall_granular_results} in the Appendix presents a complete list of code perturbations used in this paper.

\emph{Docstring Perturbations} rewrite natural language in docstrings and comments, including edits like substituting synonyms
(Figure~\ref{fig: ptb_docstring_funcname_exp}).  % shows an example of the SynonymSubstitution perturbation.
\emph{Function-Name Perturbations} refactor some function names, e.g., changing from snake\_case to camelCase (Figure~\ref{fig: ptb_docstring_funcname_exp}). % shows an example of the CamelCase perturbation.
\emph{Code-Syntax Perturbations} apply perturbations related to code syntax, involving changes like renaming variables (Figure~\ref{fig: ptb_syntax_format_exp}). % shows an example of the VarRenamer perturbation.
\emph{Code-Format Perturbations} change the code snippets' format, e.g., splitting a line into two (Figure~\ref{fig: ptb_syntax_format_exp}). % shows an example of the SplitLines perturbation.

% \linbo{I'm confused by later content so back to this definition. so is each $T_i$ a perturbation and $\pi$ is a set of perturbation? or $\pi$ is a single perturbation consisting of a few transformation $T_i$?}
% \linbo{$\trans: 2^{\alphabet^*} \mapsto \alphabet^*$?}
% \yhmodify{
% We formally define a code perturbation $\perturbation = \{\trans_1, \trans_2, \ldots\}$ as a set of string transformations, 
% where $\trans: 2^{\alphabet^* \mapsto \alphabet^*}$  and $\alphabet^*$ denotes a var-length sequence over the vocabulary $\alphabet$. 
% Here, each string transformation takes input as a sequence over the vocabulary $\alphabet$ and returns a perturbed sequence.
% Each string transformation in $\perturbation$ specifies the choice of different positions and replacements where the perturbation can take place.
% }
% {
A code perturbation is a collection of string transformations, denoted as $\perturbation = \{\trans_1, \trans_2, \ldots\}$. 
Each transformation $\trans: \mathcal{X} \mapsto \mathcal{X}$ operates on a token sequence from the input domain $\mathcal{X}$, altering them to produce a perturbed sequence. 
Within $\perturbation$, each transformation specifies different positions and replacements for perturbation.
% }

\begin{example}
    % \yhmodify{In Figure~\ref{fig: ori_partial_exp}, the VarRenamer perturbation contains an infinite set of code transformations that specify, 1) which variable name to change, and 2) the new variable name.}
    % {
    The VarRenamer perturbation, shown in Figure~\ref{fig: ori_partial_exp}, is a code perturbation $\pi$.
    It contains an infinite set of string transformations that specify 1) which variable name to change and 2) the new variable name.
    % }
    The former contains two choices: \code{n} and \code{i}.
    And the latter contains infinite choices of valid variable names. 
    % \linbo{Now the statement is much easier to understand! Nice work!}
\end{example}

% \technique is designed to be general and applicable not only to existing ReCode perturbations but also to extend new perturbations. 
We introduce two categories, \emph{context-free} and \emph{context-sensitive} perturbations, which serve as a high-level interface for robust training. % within our framework, \technique. 
%The unique attributes of each category are detailed in subsequent sections
We formalize the distinctive characteristics of these two categories in the following sections. 

% We introduce two abstraction categories, \emph{context-free} and \emph{context-sensitive} perturbations, which serve as a high-level interface for perturbed dataset generation and robust training in our robust training framework, \technique.
% We formalize the distinctive characteristics of these two categories in the following sections. 

\subsubsection{Context-Free Perturbations}
A code perturbation $\perturbation$ is a \emph{context-free} perturbation if all perturbed prompts generated by $\perturbation$ should not affect the ground-truth completion.
Formally, for all $\trans \in \perturbation$, the concatenation of the prompt perturbed by $\trans$ and the ground-truth completion remains a correct function:
\vspace{-0.5em}
\begin{align}
    \forall \trans \in \perturbation, \cor{\trans(\bfx_{:i})}{\bfx_{i:}}\label{eq: context-free-def}
\end{align}

\vspace{-0.5em}
\begin{example}
    In Figure~\ref{fig: ptb_docstring_funcname_exp}, the SynonymSubstitution perturbation in the docstring will not affect the ground-truth completion. 
\end{example}

\subsubsection{Context-Sensitive Perturbations}
A code perturbation $\perturbation$ is a \emph{context-sensitive} perturbation if any perturbed prompt generated by $\perturbation$ results in \emph{coherent} changes to the ground-truth completion. 
Formally, for all $\trans \in \perturbation$, the concatenation of the prompt perturbed by $\trans$ and its ground-truth completion perturbed correspondingly is a correct function, while the concatenation of the perturbed prompt and the original completion is not.
\vspace{-0.5em}

\begin{small}
\begin{align}
    \forall \trans \in \perturbation, \cor{\trans(\bfx_{:i})}{\trans(\bfx_{i:})} \wedge \neg \cor{\trans(\bfx_{:i})}{\bfx_{i:}} \label{eq: context-sensitive-def}
\end{align}
\end{small}

\vspace{-0.5em}
\begin{example}
    In Figure~\ref{fig: ptb_syntax_format_exp}, the VarRenamer perturbation requires the ground-truth completion to change \textbf{coherently} because all the variable \code{i} should be renamed to \code{b}. 
\end{example}

\subsection{Robustness of Code Generation Models}
% CLM is formulated as a next-token prediction task. And we want the model $f$ to predict the ground-truth completion to the prompt, i.e., $f(\bfx_{:i})=\bfx_{i:}$. 
To define model robustness, we say a model $f$ is robust to a perturbation $\perturbation$, if
\vspace{-0.5em}
\begin{align}
    \forall \trans \in \perturbation, \cor{\trans(\bfx_{:i})}{f(\trans(\bfx_{:i}))} \label{eq: robustness}
\end{align}

\vspace{-0.5em}
The robustness against context-free perturbations is similar to the traditional robustness definition, in which the perturbation should not change the model results (Eq~\ref{eq: context-free-def}). 
However, the robustness against context-sensitive perturbations differs from the traditional definition, as the context-sensitive robustness requires the model output to change coherently with the perturbed prompt (Eq~\ref{eq: context-sensitive-def}).

%% file: sections/approach.tex
\section{\technique}
\label{sec: framework}
% \begin{figure*}
%     \centering
%     \includegraphics[width=0.9\textwidth]{imgs/overview.png}
%     \caption{Overview of our approach. In the paired token sequences, white segments denote unperturbed segments, blue segments are perturbed by context-free perturbations, and the orange segments are perturbed by context-sensitive perturbations. \todored{draw a better one. The masking perturbed tokens approach has been included in the batch augmentation.}}
%     \label{fig:overview}
% \end{figure*}

% We provide an overview of our robust training framework in Figure~\ref{fig:overview}. 
% For a code snippet in the original training set, we use ReCode to randomly choose and apply one or more code perturbations to get a perturbed code snippet. 
% \technique enriches the training set using a paired dataset generation module outlined in Section~\ref{sec: approach_gendata}.
% It first tokenizes the code snippets before and after perturbations into a pair of the original and the perturbed token sequences, forming a paired dataset. 
% Furthermore, each token segment in the original sequence is paired with its counterpart in the corresponding perturbed sequence. 
% In Section~\ref{sec: approach_robust_train}, we design different robust training approaches tailored to code generation models utilizing the paired dataset.

\technique enhances the training set with a paired dataset generation method (Section~\ref{sec: approach_gendata}). 
% This process involves tokenizing code snippets before and after perturbations into pairs of original and perturbed token sequences, thereby creating a paired dataset. Each token in the original sequence is matched with its equivalent in the perturbed sequence. 
Section~\ref{sec: approach_robust_train} outlines our robust training strategies. % specifically devised for code generation models using this paired dataset.

\subsection{Paired Dataset Generation Method}
\label{sec: approach_gendata}
% The paired dataset generation approach provides two levels of pairing, \emph{example-level} pairing and \emph{sequence-level} pairing. 
% At the example level, each original training example is paired with its corresponding perturbed example. 
% At the sequence level, finer granularity is achieved by pairing each token segment in the original training example with its counterpart in the corresponding perturbed example. 
% Both levels of paring are crucial for the robust training approaches, which will be introduced in Section
The paired dataset generation method offers two levels of pairings: \emph{example-level} and \emph{sequence-level}. 
Example-level pairing matches each original training example with its perturbed counterpart. Sequence-level pairing provides more detail by matching each token segment from the original example to its equivalent in the perturbed example. These pairing mechanisms are essential for the robust training strategies discussed in Section~\ref{sec: approach_robust_train}.

\paragraph{Example-level Pairing}
Given a training set $D=\{\trpoint{\bfx}{j}\}_{j=1}^m$, where each training example is a code snippet, the paired dataset generation method returns a set of paired of training samples 
$\{(\trpoint{\bfx}{j}, \trpoint{\pt{\bfx}}{j})\}$, where $\trpoint{\pt{\bfx}}{j}$ is the code snippet perturbed by some code perturbations.
% $(D=\{\trpoint{\bfx}{j}\}_{j=1}^m, \pt{D}=\{\trpoint{\pt{\bfx}}{j}\}_{j=1}^m)$. 
% The training examples in the paired dataset are paired, i.e., $\trpoint{\bfx}{j}$ is the original code snippet, and $\trpoint{\pt{\bfx}}{j}$ is the code snippet perturbed by some code perturbations.
% We use ReCode to randomly choose and apply one or more code perturbations to a code snippet to get the perturbed code snippet. 
To obtain $\trpoint{\pt{\bfx}}{j}$, we first randomly choose $t$ code perturbations $\{\perturbation_1, \ldots, \perturbation_t\}$ in ReCode.
Then, we randomly choose one string transformation from each code perturbation and apply it to the original code snippet $\bfx$, 
\vspace{-0.5em}

\begin{small}
\begin{align}
    \pt{\bfx} = T_t(T_{t-1}(\ldots T_1(\bfx)\ldots)), \quad T_i \in \perturbation_i, \forall 1 \le i \le t. \label{eq: train_ptb}
\end{align}
\end{small}

\vspace{-0.5em}
\paragraph{Sequence-level Pairing} 
Given a pair of code snippets, $(\bfx, \pt{\bfx})$, 
% where $\bfx$ is the original code snippet and $\pt{\bfx}$ is the perturbed one, 
\technique further provides finer-granularity paring for this pair. 
\begin{example}
\label{exp: listing}
    Consider following pairs of tokens,
\begin{lstlisting}[language=prompt]
Index:      0 1 2 3 4 5 6 7 8 9
Original:   A @@C@@ D E F ~~G H~~ @@C@@ @@I@@
Perturbed:  A ~~B~~ @@X@@ D E F ~~Y~~ @@X@@ @@Z Z@@
\end{lstlisting}
In this example, the code perturbations perform two context-free perturbations, insert ``B'' and substitute ``G H'' with ``Y'', and two context-sensitive perturbations, substitute all ``C'' with ``X'' and substitute all ``I'' with ``Z Z''.
\end{example}

To create sequence-level pairing, we introduce a mask sequence $\mask$ ($\pt{\mask}$) for the original sequence $\bfx$ (and the perturbed one $\pt{\bfx}$, respectively). 
Each mask value indicates which kind of perturbation is applied to the corresponding token, \unptb{} for unperturbed, \freeptb{} for context-free, and \senptb{} for context-sensitive.
\begin{example}
\label{exp: mask}
   We show two masks of Example~\ref{exp: listing}.
\begin{lstlisting}[language=prompt]
Index:          0 1 2 3 4 5 6 7 8 9
Original Mask:  U @@S@@ U U U ~~F~~ ~~F~~ @@S@@ @@S@@
Perturbed Mask: U ~~F~~ @@S@@ U U U ~~F~~ @@S@@ @@S@@ @@S@@
\end{lstlisting}
\end{example}

This two-level pairing design is the key to enabling some of the robust training approaches, which will be introduced subsequently. 

\subsection{Designing Robust Training Approaches}
\label{sec: approach_robust_train}
This section introduces four robust training approaches in \technique.
% While data augmentation and batch augmentation are model-agnostic strategies, the latter two, adversarial logits pairing and contrastive learning, are specifically tailored for CLM training.

\subsubsection{Data Augmentation}
Data augmentation is a widely used approach to improve the robustness of machine learning models. A common practice is replacing a certain portion, denoted as $p$, of the original training examples with their perturbed counterparts in each training batch. 
Formally, for a training batch $\{\trpoint{\bfx}{j}\}_{j=1}^b$ and its paired perturbed batch $\{\trpoint{\pt{\bfx}}{j}\}_{j=1}^b$, the objective function is expressed as follows,
\vspace{-0.5em}

\begin{small}
\begin{align}
\lossdataaug = \sum_{j=1}^b a_j\lossclm(\trpoint{\pt{\bfx}}{j}) + (1-a_j)\lossclm(\trpoint{\bfx}{j}), \label{eq: lossdataaug}
\end{align}
\end{small}%
where  $a_j\overset{\text{i.i.d}}{\sim} \bernoulli{p}$ is a Bernoulli variable
%\linbo{independent in each draw? if yes, maybe add subscript $i=1$ to $b$ and write $a_i\overset{\text{i.i.d}}{\sim} \bernoulli{p}$} 
indicating whether the $j$-th training example will be perturbed or not. 

\paragraph{Masking Unnatural Perturbed Tokens} 
Some context-free perturbations introduce unnatural tokens, such as DeadCode Insertion adding an artificial code segment and ButterFingers introducing typos.
Referring back to the robustness property in Eq~\ref{eq: robustness}, our goal is for the model to learn to respond to these perturbations rather than to generate them. 
Learning to generate these unnatural perturbed tokens could adversely affect the original model performance, leading to issues like generating dead code or typos.
We propose masking the CLM loss of these unnatural perturbed tokens to address these issues.
% We denote $\lossclm(\bfx, \mask)$ as the CLM loss for the example $\bfx$ after masking out the unnatural perturbed tokens, formally,
We define the CLM loss for the example $\bfx$ after masking out the unnatural perturbed tokens
\vspace{-0.5em}
\begin{align*}
    \lossclm(\bfx, \mask) = -\sum_{i=1}^{|\bfx|} \mathds{1}_{\{m_i \neq \mathrm{\freeptb}\}} \log p_f(x_i \mid \bfx_{:i}),
\end{align*}
where $m_i \neq \mathrm{\freeptb}$ means that the $i$-th token is not perturbed by a context-free perturbation (see Example~\ref{exp: mask}). We design the masked data augmentation loss $\lossmdataaug$ by replacing the term $\lossclm(\pt{\bfx})$ in Eq~\ref{eq: lossdataaug} with the masked loss $\lossclm(\pt{\bfx}, \mask)$.

% \begin{small}
% \begin{align}
% \lossmdataaug = \sum_{\substack{\bfx \in \batch,\\ a \in \bernoulli{p}}} a\lossclm(\pt{\bfx}, \pt{\mask}) + (1-a)\lossclm(\bfx) \label{eq: lossmdataaug}
% \end{align}
% \end{small}%
% \linbo{same here}Note that we apply this masking method to all robust training approaches.

% \input{tables/equations}

\subsubsection{Batch Augmentation}
Batch augmentation~\citep{batchaug} duplicates a portion of training examples within the same batch with different perturbations.
It differs slightly from data augmentation, where a batch contains $p$ perturbed and $1-p$ original data. 
In contrast, batch augmentation \emph{augments} the entire batch with $p$ perturbed rather than \emph{replacing} $p$ original data with perturbed data as in data augmentation.
Given a training batch and its paired perturbed batch, the objective of batch augmentation is defined as follows,
\vspace{-0.5em}
\begin{align*}
\lossmbatchaug = \sum_{j=1}^b a_j\lossclm(\trpoint{\pt{\bfx}}{j}, \trpoint{\pt{\mask}}{j}) + \lossclm(\trpoint{\bfx}{j}) 
% \label{eq: lossmbatchaug}
\end{align*}
% Note that we also apply the masking mechanism to batch augmentation as well. 
When batch augmentation was originally proposed, its goal was not to improve the model robustness. 
We hypothesize that batch augmentation can further improve the robustness over data augmentation, as indicated in some multilingual cases~\citep{batchissue1, batchissue2}.

\subsubsection{Adversarial Logits Pairing}
\label{sec: alp}
Adversarial Logits Pairing (ALP)~\citep{ALP} improves the robustness of classification models by minimizing the KL divergence between the original input's prediction distribution and the perturbed input's prediction distribution. 
However, adapting ALP from classification models to generation models trained by CLM is challenging.
One straightforward approach is decomposing the generation task into multiple next-token prediction tasks.
However, the original and the perturbed token sequences can have different lengths due to some transformations adding or removing tokens.
This length discrepancy creates mismatches of each token's prediction between the two sequences.
% \linbo{This paragraph is to call out the challenge. I feel like it can be compressed. Not fully understood so will let Yuhao do it.}

To address this challenge, we leverage the sequence-level pairing provided by our paired dataset generation method.
We apply ALP only to the unperturbed segments of the two sequences, marked by $\mathrm{\unptb}$ in Example~\ref{exp: mask}.  
All unperturbed segments have the same length, allowing us to apply ALP to the predictions of these unperturbed tokens.
We use $\bfu$ and $\pt{\bfu}$ to denote the ordered indices for all unperturbed tokens in the original and perturbed sequences.
The ALP objective is defined as follows, 
\vspace{-0.5em}
\begin{align*}
\lossalp = \sum_{j=1}^b \sum_{i=1}^{|\trpoint{\bfu}{j}|} \KL\left(p_f(\cdot \mid \trpoint{\pt{\bfx}}{j}_{:\trpoint{\pt{u}}{j}_i})\parallel p_f(\cdot \mid \trpoint{\bfx}{j}_{:\trpoint{u}{j}_i})\right) %\label{eq: lossalp}
\end{align*}

\vspace{-0.5em}
\begin{example}
    In Example~\ref{exp: listing}, $\bfu=(0,2,3,4)$ and $\pt{\bfu}=(0,3,4,5)$.
\end{example}

\paragraph{ALP with name-Dropout (ALPD)}
We design another ALP objective (ALPD) specifically tailored to variable and function renaming among context-sensitive perturbations.
ALPD reduces the model's reliance on specific variable and function names by setting the attention masks of a portion of these names to zero. 
It can be seen as a dropout mechanism specific to entity names. 

We use $\drop{\bfx}$ to denote the input sequence after name-specific dropout.
\vspace{-1em}

\begin{small}
\begin{align*}
\lossalpdrop = \sum_{j=1}^b \sum_{i=1}^{|\trpoint{\bfu}{j}|} &\KL\left(p_f(\cdot \mid \drop{\trpoint{\bfx}{j}}_{:\trpoint{u}{j}_i})\parallel p_f(\cdot \mid \trpoint{\bfx}{j}_{:\trpoint{u}{j}_i})\right) \nonumber \\ 
+ &\KL\left(p_f(\cdot \mid \drop{\trpoint{\pt{\bfx}}{j}}_{:\trpoint{\pt{u}}{j}_i})\parallel p_f(\cdot \mid \trpoint{\pt{\bfx}}{j}_{:\trpoint{\pt{u}}{j}_i})\right) 
%\label{eq: lossalpdrop}
\end{align*}
\end{small}%
$\lossalpdrop$ sums two KL divergence losses: the first over the original sequence after and before dropout, and the second over the perturbed sequence.

\subsubsection{Contrastive Learning}
Contrastive learning (CL) maximizes the cosine similarity between positive (similar) pairs and minimizes the distance between negative (dissimilar) pairs.
The granularity of pairs leads to different designs of CL objectives. 
This section introduces three designs of CL objectives tailored to CLM.
ContraSeq and ContraToken, inspired by ContraCLM~\citep{contraclm}, focus on sequences and tokens, respectively.
A novel ContraName objective focuses on variable and function names. 

\paragraph{ContraSeq} The ContraSeq objective operates at the sequence level, where each pair consists of summarizations of two input sequences. We note that ContraSeq is also adopted in ContraBERT~\citep{contrabert} and ContraCode~\citep{contracode} for improving the robustness of the encoder model trained on masked language modeling (MLM).
Since CLM does not have the [CLS] token used in MLM, we compute the average of the hidden states in the last layer as the summarization.

Given a batch $\batch=\{\bfh_1, \ldots, \bfh_b, \pt{\bfh}_1, \ldots, \pt{\bfh}_b\}$ with $2b$ summarizations of original and perturbed sequences,
ContraSeq treats the $b$ corresponding original and perturb pairs, i.e., $(\bfh_i, \pt{\bfh}_i)$, as positive pairs and other pairs in the batch as negatives. 
% We provide the definition of ContraSeq ($\losscontraseq$) in Appendix~\ref{appendix: cl_objectives}.

Denoting the temperature
hyper-parameter as $\tau$ and cosine similarity as $\diamond$, we define the ContraSeq objective as follows,
\begin{align*}
\losscontraseq = \sum_{j=1}^b g(\trpoint{\bfh}{j}, \trpoint{\pt{\bfh}}{j}, B) + g(\trpoint{\pt{\bfh}}{j}, \trpoint{\bfh}{j}, B), %\label{eq: losscontraseq}
\end{align*}
where $g(x, y, B)$ is defined as 
\begin{small}
\begin{align*}
    g(x, y, B) = -\log{\frac{\exp(x \diamond y/\tau)}{\sum_{h\in B} \exp(x \diamond h/\tau) - \exp(1/\tau)}}.
    %\label{eq: the_g}
\end{align*}
\end{small}

ContraSeq represents the coarsest granularity among the three objectives.
While ContraSeq is shown to be effective for MLM, it may not fully cater to CLM, which predicts the next token for each prefix and involves discriminating representations at finer levels, e.g., tokens and names.
Additionally, ContraSeq poses scalability challenges, as it demands a large batch size to compute a meaningful InfoNCE loss. This challenge restricts ContraSeq's feasibility to large language models.

\paragraph{ContraToken} The ContraToken objective operates at the token level, providing a finer granularity than ContraSeq.
ContraToken aims to discriminate the representation of each prefix.
However, as we mentioned in Section~\ref{sec: alp}, directly treating $\bfx_{:i}$ and $\pt{\bfx}_{:i}$ as a positive pair does not work due to the potential sequence length difference between original and perturbed sequences.
To address this, ContraToken considers two prefixes ending at the same unperturbed token as a positive pair, with other prefix pairs designated as negatives. 
% Appendix~\ref{appendix: cl_objectives} provides the definition of ContraToken ($\losscontratoken$).

For the $j$-th training example, let $\trpoint{\bfh}{j}_i$ denote the representation of the prefix up to the $i$-th token, and $\trpoint{u}{j}_i$ denote the index of the $i$-th unperturbed token. We define ContraToken objective as:
\begin{small}
\begin{align*}
\losscontratoken = \sum_{j=1}^b\sum_{i=1}^{|\trpoint{\bfu}{j}|} g(\trpoint{\bfh}{j}_{\trpoint{u}{j}_i}, \trpoint{\pt{\bfh}}{j}_{\trpoint{\pt{u}}{j}_{i}}, \trpoint{H}{j})
+
g(\trpoint{\pt{\bfh}}{j}_{\trpoint{\pt{u}}{j}_{i}}, \trpoint{\bfh}{j}_{\trpoint{u}{j}_i}, \trpoint{H}{j})
%\label{eq: losscontratoken}
\end{align*}
\end{small}%
where $\trpoint{H}{j}$ contains all the representations of prefixes ending at unperturbed tokens for the $j$-th training example, i.e., $\trpoint{H}{j}=\{\trpoint{\bfh}{j}_{\trpoint{u}{j}_{1}},\ldots, \trpoint{\bfh}{j}_{\trpoint{u}{j}_{|\trpoint{\bfu}{j}|}},\trpoint{\pt{\bfh}}{j}_{\trpoint{\pt{u}}{j}_{1}},\ldots, \trpoint{\pt{\bfh}}{j}_{\trpoint{\pt{u}}{j}_{|\trpoint{\pt{\bfu}}{j}|}}\}$.

\paragraph{ContraName}
We design a novel name-level CL objective, ContraName, to address variable and function renaming in context-sensitive perturbations. 
It aims to enhance the discrimination of representations of variable and function names.

In ContraName, we group representations of variables or functions according to their names. 
For a name spanning multiple tokens, we use the average of these tokens as its representation.
ContraName treats representations within the same group as positive pairs and those across different groups as negative pairs.
Notice that the negative pairs in ContraName have explicit semantic differences, i.e., different names should yield different representations. 
This explicit semantic difference of negative pairs has been shown to improve the effectiveness of CL~\citep{hardnegative}.
% Appendix~\ref{appendix: cl_objectives} provides the definition of ContraToken ($\losscontraname$).

Suppose in the sequence $\bfx$, we identify $g$ groups of name representations $G_1, G_2, \ldots, G_g$, with $G=\bigcup
_{i=1}^g G_i$ being their union. We define the ContraName objective on the input $\bfx$ as follows,
\vspace{-0.5em}

\begin{small}
\begin{align*}
\losscontraname(\bfx) = -\log\left( \dfrac{\sum_{i=1}^g \sum_{\bfh, \bfh' \in G_i} \exp(\bfh \diamond \bfh' / \tau)}{\sum_{\bfh, \bfh' \in G} \exp(\bfh \diamond \bfh' / \tau)}  \right)
\end{align*}
\end{small}

\begin{example}
Consider the original example in Example~\ref{exp: listing} with $G_1 = \{\bfh_1, \bfh_7\}$ and $G_2=\{\bfh_8\}$.
The perturbed example contains $\pt{G_1} = \{\pt{\bfh}_2, \pt{\bfh}_7\}$ and $\pt{G_2}=\{\frac{\pt{\bfh}_8 + \pt{\bfh}_9}{2}\}$.
\end{example}

The final ContraName objective is the sum of losses over the original sequences and their perturbed counterparts.
\begin{align}
\losscontraname = \sum_{j=1}^m \losscontraname(\trpoint{\bfx}{j}) +  \losscontraname(\trpoint{\pt{\bfx}}{j})  \label{eq: losscontraname}
\end{align}

%% file: sections/evaluation.tex
\section{Evaluation}

\input{tables/main_results}
\subsection{General Experimental Setup}
% \subsubsection{Models, Datasets, and Benchmarks}

\paragraph{Models} We use different robust training approaches to fine-tune different sizes of mono-lingual CodeGen models~\citep{codegen} targeting at Python: CodeGen-6B, 2B, and 350M.
We provide fine-tuning settings in Appendix~\ref{appendix: finetune}.

\paragraph{Datasets and Benchmarks} We use the Stack dataset v1.2~\citep{TheStack} as our original training dataset. 
We uses ReCode~\citep{recode} to augment the dataset by introducing different code perturbations. 
For all experiments, we set $p=25\%$ and $t=2$ (Eq~\ref{eq: train_ptb}), i.e., we apply at most two perturbations to each original code snippet.

To evaluate the model robustness, we use the ReCode benchmark, which is based on HumanEval~\citep{humaneval} and MBPP~\citep{mbpp} with a total of 1138 (164 + 974) problems.
The docstring and function-name classes are perturbed based on the original prompt.
The code-syntax and code-format classes are perturbed based on modified prompts, which are appended with half of the ground truth completion. 
% \todored{Discuss The License For Artifacts}

\paragraph{Metrics}
We use the following three metrics to assess the model performance. 

\noindent \textit{\npsk.} We use Pass@k following \citet{humaneval} to assess the nominal code generation performance. 
We name the Pass@k on unperturbed data as Nominal Pass@k (NP@k).
% For each problem in the benchmark, we draw $n$ samples from the model's output distribution.
% The NP@k metric approximates the probability of any $k$ samples passes all the test case, if we randomly choose $k$ samples out of these $n$ samples.
% In this section, we report all metrics with $n=5$ due to the high cost of evaluation. 
This metric approximates the probability of any $k$ samples passes all the test case, if we randomly choose $k$ samples out of $n$ samples generated by the model for each problem. 
We use $n=5$ because the difference between $n=10$ and $n=100$ is already small as demonstrated in Recode~\citep{recode}.
% In this section, we report all metrics with $n=5$ due to computational constraints.
% \myshang{not sure if we should say high cost of evaluation. Maybe high cost of sampling is more appropriate?}
% \yh{We could also quote the results in recode, which shows that the difference between $n=5$ and $n=100$ is not large.}
% \todored{We report the main results of CodeGen-350M using $n=100$ in the Appendix.}

\noindent \textit{\rpsk.} To evaluate the robustness of models, we use the Robust Pass$_s$@k (RP$_{s}$@k). It measures the worst-case Pass@k on $s$ perturbed variance for each perturbation type and each sample. Here, we use $s=10$ to harden the robustness gain for training and differentiate performance gaps.
% \todored{@Shiqi explain \rpsk}

\noindent \textit{\rdrop.} We report Robust Drop\%. It measures the percentage drop from Robustness Pass@k (RP$_{s}$@k) from Nominal Pass@k (NP@k), indicating the relative robustness changes given perturbations. Lower \rdrop means better robustness.
% \paragraph{\rperct} The \rperct{} presents how much of the correct generations are robust in terms of Pass@k, i.e., \rperct=$\frac{\text{\rpsk}}{\text{\npsk}}$. \yh{need to decide whether to use \rperct or 1 - \rperct, i.e., the RD in Recode paper.}

\subsection{Effectiveness of Proposed Approaches}
\label{sec: main_res}

\begin{tcolorbox}[sharp corners, boxsep=0pt,left=3pt,right=3pt,top=2pt,bottom=2pt]
\textbf{Summary of Results:}
Our approach significantly enhances the robustness of code generation models, surpassing the results of data augmentation. Notably, our approach exhibits the most substantial improvement in robustness against Syntax perturbations.
\end{tcolorbox}

Table~\ref{tab:main_results} summarizes the robust evaluation results for CodeGen models. 
We use $\lossclm$ to denote the baseline method fine-tuning on the original stack dataset (unseen by CodeGen models) without any robust training approaches. 
Comparing $\lossclm$ and the original model (Ori), we find that fine-tuning on unseen data can already improve model robustness on the Docstring, Function, and Format perturbations, except the Syntax perturbation.

% When averaging across four perturbation classes, our approach demonstrates significant improvements in \rpsk (\rdrop)—7.76 (22.86), 6.49 (18.38), and 4.99 (18.90) for CodeGen-6B, CodeGen-2B, and CodeGen-350M, respectively, compared to the baseline $\lossclm$. In contrast, data augmentation achieves sub-optimal results with improvements of 7.91 (19.47), 6.24 (16.98), and 3.43 (12.38).
When averaging across four perturbation classes, our approach demonstrates significant improvements in \rpsk—9.37, 6.49, and 4.99 for CodeGen-6B, CodeGen-2B, and CodeGen-350M, respectively, compared to the baseline $\lossclm$. In contrast, data augmentation achieves sub-optimal results with improvements of 7.87, 6.24, and 3.12.
% We note that data augmentation achieves better \rpsk on CodeGen-6B than our approach because the \npsk of our approach is much lower.
% Moreover, our approach even improves the nominal pass rate on \todored{xxx models}.

Averaging over all three models, our approach enhances \rpsk by 3.29, 0.88, 17.94, and 5.68 for Docstring, Function, Syntax, and Format perturbations, respectively, compared to the baseline $\lossclm$.
Surprisingly, our approach exhibits the most substantial improvement in robustness against Syntax perturbations. This emphasis on strengthening robustness to Syntax perturbations is crucial for ensuring the reliability of code models in handling diverse syntactic variations.

We conducted statistical analyses using paired-t tests to compare our approach with baseline $\lossclm$ and data augmentation $\lossdataaug$ across four perturbation classes.
Our approach significantly outperforms the baseline $\lossclm$ with $p<0.05$ on all perturbation classes and all models with exceptions of function-name perturbations on CodeGen-6B and CodeGen-2B.
We hypothesize that the less pronounced results on function-name perturbations are due to the imbalanced perturbed data, as the percentages of function-name perturbations are much smaller compared to other perturbation types. 
When comparing our approach with data augmentation $\lossdataaug$ (shown in Table~\ref{tab:main_results}), we found that our approach significantly outperforms $\lossdataaug$ with $p<0.01$ on six cases, while $\lossdataaug$ outperforms our approach with $p<0.05$ on one case.

\subsection{Ablation Studies}
\label{sec: ablations}

\begin{tcolorbox}[sharp corners, boxsep=0pt,left=3pt,right=3pt,top=2pt,bottom=2pt]
\textbf{Summary of Results:}
The ablation studies confirm the effectiveness of masked batch augmentation, ALP, and ALPD.
ContraSeq provides negligible improvements compared to the baseline ($\lossclm$).
ContraToken and ContraName yield mixed results in different settings.
\end{tcolorbox}

This section presents the ablation results of different approaches outlined in Section~\ref{sec: approach_robust_train} applied to the CodeGen-350M model.
We conduct our experiments in two settings.
The first setting (Table~\ref{tab:ablation_results_free}) focuses on context-free perturbations, applying the original data augmentation ($\lossdataaug$) loss to context-sensitive perturbations while varying different approaches for the context-free perturbations. 
In the second setting (Table~\ref{tab:ablation_results_sensitive}), we vary approaches for context-sensitive perturbations while maintaining the $\lossdataaug$ loss for context-free perturbations.
We report the overall average of \npsk, \rpsk, and \rdrop across four perturbation classes, with more details reported in Appendix~\ref{sec: appendix_ablation_detail}.

\input{tables/ablation}

% \subsubsection{Ablation Results on Context-Free Perturbations}

\paragraph{Effectiveness on Masked Batch Augmentation} 
The masked batch augmentation loss $\lossmbatchaug$ consists of two components: (1) a masking mechanism that masks unnatural perturbed tokens and (2) batch augmentation.
Comparing the results of masked data augmentation ($\lossmdataaug$) and data augmentation ($\lossdataaug$) in Table~\ref{tab:ablation_results_free} validates the effectiveness of the masking mechanism because $\lossmdataaug$ achieves better \rpsk and \rdrop than $\lossdataaug$.
To assess the effectiveness of batch augmentation, we cannot directly compare the results of $\lossmdataaug$ ($[2]$, Table~\ref{tab:ablation_results_free}) and $\lossmbatchaug$ ($[4]$, Table~\ref{tab:ablation_results_free}) because $\lossmdataaug$ is trained on $p=25\%$ perturbed data, while $\lossmbatchaug$ is trained on $\frac{p}{1+p}=20\%$ perturbed data.
For a fair comparison, we train $\lossdataaug$ with $p=20\%$ perturbed data and report the result at $[3]$ in Table~\ref{tab:ablation_results_free}.
Comparing the results of $\lossmdataaug(p=20\%)$ and $\lossmbatchaug$ confirms the effectiveness of batch augmentation because $\lossmbatchaug$ achieves better \npsk, \rpsk, and \rdrop than $\lossmdataaug(p=20\%)$.

\paragraph{Effectiveness of ALP and ALPD}
ALP and ALPD are shown to be effective because $\lossalp$ and $\lossalpdrop$ both improve the \rpsk and \rdrop ($[5]$ vs $[4]$ in Table~\ref{tab:ablation_results_free} and $[1]$ vs $[2]$ in Table~\ref{tab:ablation_results_sensitive}).
We further investigate different designs of ALP and ALPD in Appendix~\ref{sec: appendix_ablation_detail}.

% \subsubsection{Ablation Results on Context-Sensitive Perturbations}

\paragraph{Discussion on Contrastive Learning Objectives}
ContraSeq only provides negligible improvements, as evidenced by $[6]$ vs $[0]$, and $[8]$ vs $[5]$ in Table~\ref{tab:ablation_results_free}.
ContraToken behaves differently in two ablation experiment settings. 
In the context-free perturbation experiment (Table~\ref{tab:ablation_results_free}), ContraToken improves the \rdrop but negatively impacts the \npsk and \rpsk ($[7]$ vs $[5]$). 
Conversely, ContraToken hurts all metrics ($[5]$ vs $[4]$) in the context-sensitive perturbation experiment (Table~\ref{tab:ablation_results_sensitive}). 
Adding ContraName to the masked batch augmentation loss improves the \rdrop and \rpsk ($[3]$ vs $[1]$).
% However, adding ContraName to $\lossmbatchaug + \lossalpdrop$ hurts all metrics ($[4]$ vs $[2]$). 

%% file: tables/main_results.tex
\begin{table*}
\centering
\resizebox{0.95\linewidth}{!}{
\small
\setlength{\tabcolsep}{2pt}
\renewcommand{\arraystretch}{1} 
\begin{tabular}{ll rr rr rr rr rrr}
\toprule
\multicolumn{2}{c}{\multirow{2}{*}{Model \& Methods}}  & \multicolumn{2}{c}{Docstring} & 
\multicolumn{2}{c}{Function} & \multicolumn{2}{c}{Syntax} & \multicolumn{2}{c}{Format} & \multicolumn{3}{c}{Overall Average} \\ 
\cmidrule(lr){3-4} \cmidrule(lr){5-6} \cmidrule(lr){7-8} \cmidrule(lr){9-10}  \cmidrule(lr){11-13} 
& & \npsk & \rpsk & \npsk & \rpsk & \npsk & \rpsk & \npsk & \rpsk & \npsk & \rpsk & \rdrop  \\

\midrule

% \multirow{3}{*}{StarCoder} 
% & Ori($\lossclm$) & 42.34 & 17.61 & 42.34 & 17.50 & 59.74 & 2.48 & 59.74 & 35.59 & 51.04 & 18.30 & 35.85\\
% & $\lossdataaug$ & 43.95 & 26.66 & 43.95 & 27.75 & 58.58 & 32.21 & 58.58 & 37.79 & 51.27 & 31.10 & 60.66\\
% & \cellcolor{gray!15} Ours &  \\
% \midrule

\multirow{4}{*}{CodeGen-6B}
& Ori & 35.96 & 12.83 & 35.96 & 14.36 & 52.72 & 2.20 & 52.72 & 25.47 & 44.34 & 13.71 & 69.08\\
& $\lossclm$ & 40.07 & 20.21 & 40.07 & 22.18 & 54.91 & 2.58 & 54.91 & 35.80 & \textbf{47.27} &  20.19 & 57.29\\
& $\lossdataaug$ & 37.61 & 20.51 & 37.61 & 21.88 & 52.99 & 27.66 &  52.99 & 42.18 & 45.30 & 28.06 & 38.06\\
& \cellcolor{gray!15} Ours & \cellcolor{gray!15} 37.91 & \cellcolor{gray!15} \textbf{$^{**}$23.13} & \cellcolor{gray!15} 37.91 & \cellcolor{gray!15} \textbf{22.53} & \cellcolor{gray!15} 53.16 & \cellcolor{gray!15} \textbf{27.70} & \cellcolor{gray!15} 53.16 & \cellcolor{gray!15} \textbf{$^{**}$44.87} & \cellcolor{gray!15} 45.54 & \cellcolor{gray!15} \textbf{29.56} & \cellcolor{gray!15} \textbf{35.09} \\
% & $\lossclm$ & 39.68 & 19.47 & 39.68 & \textbf{22.20} & 54.48 & 2.64 & 54.48 & 35.89 & \textbf{47.08} &  20.05 & 57.41\\
% & $\lossdataaug$ & 36.91 & 20.62 & 36.91 & 21.62 & 53.20 & 27.35 &  53.20 & \textbf{42.25} & 45.05 & \textbf{27.96} & 37.94\\
% & \cellcolor{gray!15} Ours & \cellcolor{gray!15} 35.06 & \cellcolor{gray!15} \textbf{20.90} & \cellcolor{gray!15} 35.06 & \cellcolor{gray!15} 21.07 & \cellcolor{gray!15} 51.92 & \cellcolor{gray!15} \textbf{27.43} & \cellcolor{gray!15} 51.92 & \cellcolor{gray!15} 41.83 & \cellcolor{gray!15} 42.49 & \cellcolor{gray!15} 27.81 & \cellcolor{gray!15} \textbf{34.55} \\
\midrule

\multirow{4}{*}{CodeGen-2B} 
& Ori & 31.27 & 11.04 & 31.27 & 9.75 & 44.82 & 1.63 & 44.82 & 24.45 & 38.05 & 11.72 & 69.20 \\
& $\lossclm$ & 32.99 & 15.78 & 32.99 & 16.41 & 46.22 & 2.43 & 46.22 & 32.00 & \textbf{39.61} & 16.66 & 57.94 \\
& $\lossdataaug$ & 31.62 & 17.62 & 31.62 & 16.61 & 45.96 & $^{*}$\textbf{22.86} & 45.96 & \textbf{34.50} & 38.79 & 22.90 & 40.96 \\
& \cellcolor{gray!15} Ours & \cellcolor{gray!15} 31.56 & \cellcolor{gray!15} \textbf{$^{**}$19.21} & \cellcolor{gray!15} 31.56 & \cellcolor{gray!15} \textbf{17.12} & \cellcolor{gray!15} 45.04 & \cellcolor{gray!15} 21.92 & \cellcolor{gray!15} 45.04 & \cellcolor{gray!15} 34.36 & \cellcolor{gray!15} 38.30 & \cellcolor{gray!15} \textbf{23.15} & \cellcolor{gray!15} \textbf{39.56} \\
\midrule

\multirow{4}{*}{CodeGen-350M} 
& Ori & 17.10 & 3.57 & 17.10 & 3.06 & 26.75 & 1.11 & 26.75 & 9.54 & 21.93 & 4.32 & 80.30\\
& $\lossclm$ & 18.10 & 6.19 & 18.10 & 6.47 & 29.24 & 1.46 & 29.24 & 15.96 & 23.67 & 7.52 & 68.23 \\
& $\lossdataaug$ & 18.10 & 7.45 & 18.10 & 7.94 & 30.11 & 8.59 & 30.11 & 18.58 & 24.10 & 10.64 & 55.85 \\
% $\lossdataaug$ & 18.28 & 7.93 & 18.28 & 7.41 & 30.56 & 9.35 & 30.56 & 19.12 & 24.42 & 10.95 & 44.84 \\
& \cellcolor{gray!15} Ours & 
\cellcolor{gray!15} 18.33 & 
\cellcolor{gray!15} \textbf{$^{**}$9.72} & \cellcolor{gray!15} 18.33 & \cellcolor{gray!15} \textbf{8.05} & \cellcolor{gray!15} 31.04 & \cellcolor{gray!15} \textbf{$^{**}$10.67} & \cellcolor{gray!15} 31.04 & \cellcolor{gray!15} \textbf{$^{**}$21.58} & \cellcolor{gray!15} \textbf{24.69} & \cellcolor{gray!15} \textbf{12.51} & \cellcolor{gray!15} \textbf{49.33}\\

\bottomrule
\end{tabular}
}
\vspace{-0.5em}
\caption{Robust evaluation of CodeGen-6B, 2B, 350M on the ReCode benchmark. Our approach combines $\lossmbatchaug$, $\lossalp$, and $\lossalpdrop$. We show the statistical significance between our approach and $\lossdataaug$ using the paired-t test with $^*$ denoting $p<0.05$ and $^{**}$ denoting $p<0.01$. \npsk and \rpsk are higher the better. \rdrop is lower the better.}
    \label{tab:main_results}
    \vspace{-1em}
\end{table*}

%% file: tables/ablation.tex
\begin{table}[t]
\centering
\resizebox{0.9\columnwidth}{!}{
\small
\setlength{\tabcolsep}{2pt}
\renewcommand{\arraystretch}{1} 
\begin{tabular}{l rrr}
\toprule
\multirow{2}{*}{Methods} & \multicolumn{3}{c}{Overall Average} \\ 
\cmidrule(lr){2-4} 
& \npsk & \rpsk & \rdrop  \\
\midrule
$[0]:\lossclm$ & 23.67 & 7.52 & 68.23\\
% $\lossdataaug$ & 18.28 & 7.93 & 18.28 & 7.41 & 30.56 & 9.35 & 30.56 & 19.12 & 24.42 & 10.95 & 44.84 \\
$[1]:\lossdataaug$ & 24.10 & 10.64 & 55.85 \\
$[2]:\lossmdataaug$ & 23.77 & 11.10 & 53.30\\
$[3]:\lossmdataaug (p=20\%)$  & 24.60 & 10.48 & 57.40\\
$[4]:\lossmbatchaug$ & 24.90 & 11.01 & 55.78\\
\rowcolor{gray!15}$[5]:\lossmbatchaug + \lossalp$ &  24.67 & 11.48 & 53.47 \\
$[6]:\lossclm + \losscontraseq$ & 23.80 & 7.79 & 67.27 \\
$[7]:[5] + \losscontratoken$ & 23.52 & 11.18 & 52.47 \\
$[8]:[5] + \losscontraseq$ & 24.64 & 11.51 & 53.29 \\
\bottomrule
\end{tabular}
}
\vspace{-0.5em}
    \caption{Ablation results on \emph{context-free perturbations}.}
    \label{tab:ablation_results_free}
     \vspace{-1em}
\end{table}

\begin{table}[t]
\centering
\resizebox{0.9\columnwidth}{!}{
\small
\setlength{\tabcolsep}{2pt}
\renewcommand{\arraystretch}{1} 
\begin{tabular}{l rrr}
\toprule
\multirow{2}{*}{Methods} & \multicolumn{3}{c}{Overall Average} \\ 
\cmidrule(lr){2-4} 
& \npsk & \rpsk & \rdrop  \\
\midrule
$[0]:\lossclm$ & 23.67 & 7.52 & 68.23\\
$[1]:\lossmbatchaug$ & 24.83 & 10.75 & 56.71\\
\rowcolor{gray!15}$[2]:\lossmbatchaug + \lossalpdrop$ & 24.82 & 10.94 & 55.92 \\
$[3]:\lossmbatchaug + \losscontraname$ & 24.60 & 10.87 & 55.81\\
$[4]:[2] + \losscontraname$ & 24.66 & 10.80 & 56.20 \\
$[5]:[4] + \losscontratoken$ & 24.42 & 10.42 & 57.33 \\
\bottomrule
\end{tabular}
}
\vspace{-0.5em}
\caption{Ablation results on \emph{context-sensitive} perturbations.}
    \label{tab:ablation_results_sensitive}
     \vspace{-1em}
\end{table}

%% file: sections/appendix.tex
\input{tables/motivation}

\section{Experiment Setups}
\label{appendix: finetune}
% For StarCoder fine-tuning, we train with 100\% perturbed data as the model has been fine-tuned on the original dataset $D$ for two epochs~\citep{starcoder}. 
% We set batch size to $128$ and fine-tune the model for 2K steps using the AdamW optimizer and a linear schedule with 50 warmup steps and a learning rate $2\times 10^{-5}$. 
\paragraph{Experiment Environment}
All fine-tuning experiments run on a cluster of Amazon EC2 P4 Instances. 
All evaluation experiments on ReCode run on a cluster of Amazon EC2 P4 Instances and Amazon EC2 P3 Instances.

\paragraph{Fine-tuning Settings}
% For CodeGen fine-tuning, we
We train with $p=25\%$ perturbed data as CodeGen models has not been fine-tuned on the stack dataset. 
For CodeGen-2B and CodeGen-6B, we set batch size to $256$ and fine-tune them for 10K and 5K steps, respectively, using the AdamW optimizer and a linear schedule with 500 warmup steps and a learning rate $2\times 10^{-5}$. 
For CodeGen-350M, we set batch size to $512$ and fine-tune the model on half of the stack dataset (about 266K steps) using the FusedAdam optimizer and a linear schedule with 500 warmup steps and a learning rate $2\times 10^{-5}$. 

We treat all the objective functions proposed in this paper equally, i.e., summing them up without reweighing. 
For the temperature hyperparameter $\tau$ in contrastive learning, we set $\tau = 0.05$ for all experiments following ContraCLM. We set the dropout rate to $0.1$ for $\lossalpdrop$.

\paragraph{Training Cost of Proposed Approaches}
We apply our approach to a subset of the training data, specifically $25\%$ of the examples. For this subset, our approach requires twice as much memory as standard data augmentation because it needs to see both the perturbed and the original examples simultaneously. The rest of the training costs are the same as data augmentation.

It is important to note that for the remaining $1-25\%=75\%$ of the training data, our approach has the same training cost as standard data augmentation. Considering the benefits of improved model robustness, the overall increase in training cost is relatively modest. 
Further, users can trade off the training cost and targeted robustness gain by adjusting $p$.

\paragraph{Discussion of the Licenses of Datasets}
In our paper, we employed 1) the HumanEval dataset which is distributed under the MIT license, 2) the MBPP dataset, which is under the Apache-2.0 license, 3) the CodeGen model, which is governed by the BSD-3-Clause license, and 4) the stack v1.2 dataset comprised of a collection of permissively-licensed source code.

% \section{Formal Definitions of Robust Objectives}
% \label{appendix: cl_objectives}

% \paragraph{ALP with name-Dropout (ALPD)}

% \paragraph{ContraSeq} 

% \paragraph{ContraToken} 

% \paragraph{ContraName} 

\section{Discussion on Adversarial Attacks and Adversarial Training}
Numerous adversarial attacks have targeted encoder-decoder models in code-related tasks, including classification (e.g., vulnerability prediction) and generation (e.g., code summarization).
Key methods include CODA~\citep{attack_coda}, which exploits syntactic differences for adversarial example generation; CARROT~\citep{carrot}, employing a lightweight hill climbing for optimization in attacks; and ALERT~\citep{alert}, which creates naturalness-aware attacks using pre-trained models. 

Existing work typically enhances model robustness through data augmentation and adversarial training~\citep{pgd}. 
\citet{robust_ast_hole} refine model representations by feeding only pertinent program parts to the model;
\citet{robust_delta_debugging} use curriculum learning and data augmentation with simplified programs.

Our experiments did not include adversarial training due to its significant computational overhead. For example, consider CodeAttack~\citep{code_adv5}, an adversarial attack method for pre-trained code models. 
Given an input with length $n$ and each token having $l$ possible substitutions, CodeAttack requires $O(nl)$ forward evaluations of the model in the worst-case scenario ($O(n)$ in the best-case) for each input to generate an adversarial example. 
Code generation models typically use a context-window size of $n=4096$, which means that adversarial training using CodeAttack would introduce a training overhead of at least 4096 forward evaluations for each input. 
This overhead is prohibitively expensive for fine-tuning large language models with billions of parameters.

\section{Detailed Results for Each Perturbation Type}
% Create a table to show every perturbation types
% 350M baseline vs dataaug vs combined
\input{tables/extended_results}

Table~\ref{tab:overall_granular_results} shows a detailed breakdown of robustness gain by finetuning with our approach for each perturbation type evaluated on 350M, 2B, and 6B CodeGen models.

\section{Qualitative Examples}
% picking top 4 perturbations (baseline vs combined)
In this section, we present qualitative examples to demonstrate the robustness improvements of our robust trained models. On these MBPP examples, 6B CodeGen baseline model fails to generate correct completions after applying the perturbations. Our robust trained model, on the other hand, can still successfully complete these problems. Here, we list examples for the top four perturbation types that we have achieved the most improvements (detailed numbers for each perturbation type can be found in Table~\ref{tab:overall_granular_results}).

\paragraph{DeadCode Insertion.} For this perturbation type, ReCode perturbation will insert a redundant code block including an \texttt{if condition}, a \texttt{for loop}, or a \texttt{while loop}. The models will usually be distracted by the inserted blocks, causing failure of completions. Here, in Listing~\ref{qe-deadcode}, ReCode perturbation inserts a redundant \texttt{if condition} block at line 12-13. Even though the semantic meaning barely changes after this insertion, the 6B baseline model fails to generate the correct answer.

% Deadcode
\begin{lstlisting}[language=python, caption=DeadCoderInsertion (6B CodeGen), label=qe-deadcode]
def test_distinct(data):
    """
    Write a python function to determine whether all the numbers are different from each other are not.
    >>> test_distinct([1,5,7,9])
    True
    >>> test_distinct([2,4,5,5,7,9])
    False
    >>> test_distinct([1,2,3])
    True
    """
    if len(data) == len(set(data)):
        if False:
            return True

# === completion by baseline model ===
    return False

# === completion by our model ===
        return True
    else:
        return False
\end{lstlisting}

\paragraph{NewLine Insertion.} For newline insertions, ReCode will randomly insert empty lines within prompts, and code models commonly perform poor robustness on newline insertions as reported in~\cite{recode}. In Listing~\ref{qe-newline}, 6B baseline model will generate low quality outputs if we simply insert two single empty lines at line 12 and line 15. Our models remain correct under the perturbation.

\begin{lstlisting}[language=python, caption=NewlineInsertion (6B CodeGen), label=qe-newline]
def count_first_elements(test_tup):
  """
  Write a function to find the element count that occurs before the record in the given tuple.
  >>> count_first_elements((1, 5, 7, (4, 6), 10) )
  3
  >>> count_first_elements((2, 9, (5, 7), 11) )
  2
  >>> count_first_elements((11, 15, 5, 8, (2, 3), 8) )
  4
  """

  for count, ele in enumerate(test_tup):

# === completion by baseline model ===
    if isinstance(ele, tuple):
      return count_first_elements(ele)

    else:
      return count

# === completion by our model ===
    if isinstance(ele, tuple):
      return count
  return len(test_tup)
\end{lstlisting}

\paragraph{Synonym Substitution.} To perform synonym substitution on docstrings, ReCode will randomly select words within docstrings or comments. It then replaces the selected words to their synonyms according to WordNet~\cite{miller1995wordnet}. In Listing~\ref{qe-synonym}, ``have" is replaced to be ``induce" and ``given" to ``yield". With only these two-word changes, 6B baseline model fails. In contrast, our model is much more robust to these changes.

\begin{lstlisting}[language=python, caption=SynonymSubstitution (6B CodeGen),label=qe-synonym]
def find_tuples(test_list, K):
  """
  Write a function to find tuples which induce all element divisible by k from the yield list of tuples.
  >>> find_tuples([(6, 24, 12), (7, 9, 6), (12, 18, 21)], 6)
  '[(6, 24, 12)]'
  >>> find_tuples([(5, 25, 30), (4, 2, 3), (7, 8, 9)], 5)
  '[(5, 25, 30)]'
  >>> find_tuples([(7, 9, 16), (8, 16, 4), (19, 17, 18)], 4)
  '[(8, 16, 4)]'
  """
  
  # === completion by baseline model ===
  return '[' + ','.join(map(str, filter(lambda x: x[0] % K == 0, test_list))) + ']'

  # === completion by our model ===
  return str(list(filter(lambda x: all(x[i] % K == 0 for i in range(len(x))), test_list)))

\end{lstlisting}

\paragraph{CharCase Change for Function Name.} For function name perturbation, ReCode CharCase change is to randomly replace characters in function name to their upper cases. In Listing~\ref{qe-funcchar}, ``remove\_dirty\_chars"" are replaced with ``remOve\_dIrty\_cHarS". With only four characters changed to their upper cases, 6B baseline model will fail under such perturbation. Our model demonstrates better robustness performance.

\begin{lstlisting}[language=python, caption=CharCase change for function name (6B CodeGen), label=qe-funcchar]
def remOve_dIrty_cHarS(string, second_string):
        """
        Write a function to remove characters from the first string which are present in the second string.
        >>> remOve_dIrty_cHarS("probasscurve", "pros")
        'bacuve'
        >>> remOve_dIrty_cHarS("digitalindia", "talent")
        'digiidi'
        >>> remOve_dIrty_cHarS("exoticmiles", "toxic")
        'emles'
        """

# === completion by baseline model ===
        return string.replace(second_string, "")

# === completion by our model ===
        for char in second_string:
                if char in string:
                        string = string.replace(char, "")
        return string
\end{lstlisting}

% \section{Compilation error vs assertion error}

% assertion vs non-assertion error (runtime \& compilation error)

% normalized

% robustness

% \section{unseen perturbation (or other programming language)}

\section{Detailed Results for Ablations}
\label{sec: appendix_ablation_detail}

\input{tables/ablation_full}

Tables~\ref{tab:ablation_results_free_full}~and~\ref{tab:ablation_results_sensitive_full} shows detailed comparison among different approaches across four perturbation classes.

\subsection{Different Designs of ALP}
This section compares different designs of ALP. In terms of the KL divergence loss, two approaches are considered: (1) optimizing both original and perturbed token prefixes simultaneously, i.e., bringing their output distributions closer at the same time, denoted as $\emph{Bo}$ (both sides), and (2) optimizing only the perturbed token prefix, i.e., only bringing the output distribution of the perturbed token prefix closer to the original one, denoted as $\emph{On}$ (one side).
Another aspect involves whether to optimize all prefixes or just the ones that are correctly predicted. The instance that optimizes all prefixes is named $\emph{Al}$ (all), while the one optimizing only correctly predicted prefixes is named $\emph{CO}$ (correct only). In summary, there are four different ALP designs (two by two).
Lines [9]-[12] in Table~\ref{tab:ablation_results_free_full} show that $\emph{On} + \emph{Al}$ achieves the best overall \rpsk among the four design. Therefore, we use this design throughout our experiments.

\subsection{Different Designs of ALPD}
This section compares three different designs of ALP. We conduct two additional experiments: (1) dropout of 10\% arbitrary tokens, denoted as $\emph{All}$, and (2) dropout of arbitrary tokens while following the same percentage as 10\% of variable and function names, denoted as $\emph{AllS}$ (all stratified).
Comparing line $[2]$ with lines $[6]$ and $[7]$ in Table~\ref{tab:ablation_results_sensitive_full}, we observe that $\lossalpdrop$ with $10\%$ dropout on names achieves the best overall \npsk and \rpsk. Therefore, we use this design throughout our experiments.

\subsection{Effectiveness of Combining Context-Free and Context-Sensitive Perturbations}
Based on the ablation results, we choose to use $\lossmbatchaug + \lossalp + \lossalpdrop$ for all the models in Section~\ref{sec: main_res}.
Our approach involves training on the combination of context-free and context-sensitive perturbations.
Comparing the results of our combined approach on CodeGen-350M in Table~\ref{tab:main_results} with those in Table~\ref{tab:ablation_results_free_full} line $[5]$ and in Table~\ref{tab:ablation_results_sensitive_full} line $[2]$, we observe an improvement in model robustness.
Specifically, our combined approach outperforms the other two approaches that focus solely on either context-free or context-sensitive perturbations in Docstring and Format.

%% file: tables/motivation.tex
\begin{table}
\renewcommand{\arraystretch}{1.}
\centering
\resizebox{\linewidth}{!}{
\begin{tabular}{llrrr}
\toprule
\multicolumn{2}{c}{Transformation}  &  StarCoder & WizardCoder & CodeGen16B \\

\midrule
\multirow{3}{*}{Docstring} & \npsk & 
41.27 & 53.29 & 39.23 \\ 
& \rpskf & 11.60
& 20.43 & 15.81 \\
& \rdrop & 71.89 & 61.66 & 69.70 \\
\hline

\multirow{3}{*}{Function} & \npsk & 
41.27 & 53.29 & 39.23 \\ 
& \rpskf & 15.30 & 29.06 & 26.95 \\
& \rperct & 62.93 & 45.47 & 31.30 \\
\hline

\multirow{3}{*}{Syntax} & \npsk & 59.34 & 61.09 & 56.78 \\ 
& \rpskf & 4.21 & 9.86 & 5.54 \\
& \rperct & 92.91 & 83.86 & 90.24 \\
\hline

\multirow{3}{*}{Format} & \npsk & 59.34 & 61.09 & 56.78 \\ 
& \rpskf & 23.61 & 28.13 & 39.59 \\
& \rperct & 60.21 & 53.95 & 30.27 \\
% \hline

\bottomrule
\end{tabular}
}
\caption{The ReCode~\citep{recode} robustness evaluation for SOTA public code models. \npsk shows the nominal pass@1 without perturbation; \rpskf shows the robust pass@1 under perturbation. The significant drop of Drop\% indicates unsatisfied robustness performance of these models.}
    \label{tab:motivation}
\end{table}

%% file: tables/extended_results.tex
% \usepackage{multirow}
\begin{table*}[]
\centering
\resizebox{0.9\linewidth}{!}{

    \begin{tabular}{l|l|rrr|rrr|rrr}
    \toprule
        \multicolumn{1}{c|}{\multirow{2}{*}{Categories}} & \multicolumn{1}{c|}{\multirow{2}{*}{Transformations}} & \multicolumn{3}{c|}{CodeGen 350M}                                                               & \multicolumn{3}{c|}{CodeGen 2B}                                                                 & \multicolumn{3}{c}{CodeGen 6B}                                                                 \\ \cline{3-11} 
        \multicolumn{1}{c|}{}                            & \multicolumn{1}{c|}{}                                 & \multicolumn{1}{l}{$\lossclm$} & \multicolumn{1}{l}{$\lossdataaug$} & \multicolumn{1}{l|}{Ours} & \multicolumn{1}{l}{$\lossclm$} & \multicolumn{1}{l}{$\lossdataaug$} & \multicolumn{1}{l|}{Ours} & \multicolumn{1}{l}{$\lossclm$} & \multicolumn{1}{l}{$\lossdataaug$} & \multicolumn{1}{l}{Ours} \\ \hline
    \multirow{2}{*}{Nominal}     
    & Regular                         & 18.10    & 18.10   & 18.33    & 32.99    & 31.62   & 31.56    & 40.07 & 37.61 & 37.91    \\
    & Partial                         & 29.24    & 30.11   & 31.04    & 46.22    & 45.96   & 45.04    & 54.46 & 52.99 & 53.16    \\ \hline
    \multirow{6}{*}{Docstring}   
    & BackTranslation                 & 17.35    & 17.66   & 17.79    & 31.6     & 29.86   & 30.53    & 38.91 & 36.75 & 37.45    \\
                                 & EnglishInflectionalVariation    & 10.98    & 10.98   & 12.50    & 23.95    & 22.93   & 23.99    & 28.35 & 27.21 & 29.02   \\
                                  & SynonymSubstitution\cellcolor{gray!30}              & 7.03\cellcolor{gray!30}      & 8.98\cellcolor{gray!30}     & 11.20\cellcolor{gray!30}     & 17.35\cellcolor{gray!30}     & 18.95\cellcolor{gray!30}    & 21.18\cellcolor{gray!30}     & 22.11\cellcolor{gray!30}     & 22.78\cellcolor{gray!30}    & 25.06\cellcolor{gray!30}     \\
                                 & TenseTransformationFuture       & 17.49    & 17.72   & 18.33    & 32.07    & 31.34   & 31.35    & 39.79 & 37.38 & 37.63    \\
                                 & TenseTransformationPast         & 18.12    & 18.51   & 18.63    & 32.55    & 31.21   & 31.55    & 39.40 & 37.28 & 37.70   \\
                                 & WorstCase                       & 6.19     & 7.45    & 9.72     & 15.78    & 17.07   & 19.21    & 20.21 & 20.51 & 23.13    \\ \hline
    \multirow{7}{*}{Function}    & RenameButterFinger          & 11.56    & 11.79   & 11.90    & 23.88    & 23.15   & 23.34    & 29.86 & 28.95 & 28.82    \\
                                 & RenameCamelCase             & 17.70    & 17.72   & 18.15    & 34.08    & 32.06   & 32.37    & 40.47 & 37.91 & 38.59    \\
                                 & RenameChangeChar \cellcolor{gray!30}             & 8.54\cellcolor{gray!30}      & 10.39\cellcolor{gray!30}    & 10.33\cellcolor{gray!30}     & 20.14\cellcolor{gray!30}     & 20.91\cellcolor{gray!30}    & 20.88\cellcolor{gray!30}     & 27.03\cellcolor{gray!30}     & 26.63\cellcolor{gray!30}    & 26.84\cellcolor{gray!30}     \\
                                 & RenameInflectionalVariation & 14.11    & 14.76   & 15.20    & 28.56    & 27.87   & 28.19    & 33.36 & 32.48 & 33.66    \\
                                 & RenameSwapChar              & 11.92    & 11.86   & 12.20    & 24.69    & 24.17   & 24.25    & 31.44 & 29.81 & 29.49    \\
                                 & RenameSynonymSub            & 12.07    & 12.95   & 13.04    & 24.97    & 24.50   & 24.82    & 30.14 & 29.95 & 30.47    \\
                                 & WorstCase                       & 6.47     & 7.94    & 8.05     & 16.41    & 16.61   & 17.12    & 22.18 & 21.88 & 22.53    \\ \hline
    \multirow{8}{*}{Syntax}      & DeadCodeInsertion                & 1.92     & 15.83   & 20.77    & 3.87     & 33.25   & 32.93    & 3.32 & 38.86 & 41.16    \\
                                 &  DeadCodeInsertionLast \cellcolor{gray!30}             & \cellcolor{gray!30} 9.24     & \cellcolor{gray!30} 31.55   & \cellcolor{gray!30} 32.69    & \cellcolor{gray!30} 13.90    & \cellcolor{gray!30} 48.26   & \cellcolor{gray!30} 49.47    & \cellcolor{gray!30} 14.39    & \cellcolor{gray!30} 55.13    & \cellcolor{gray!30} 55.15    \\
                                 & ForWhileTransformer             & 27.08    & 26.99   & 29.16    & 43.78    & 42.76   & 41.90    & 50.35 & 50.49 & 50.81    \\
                                 & OperandSwap                     & 27.80    & 26.91   & 29.12    & 44.50    & 43.32   & 43.15    & 51.53 & 51.46 & 51.60    \\
                                 & VarRenamerCB                    & 26.52    & 25.85   & 27.72    & 44.60    & 42.85   & 42.04    & 49.12 & 48.17 & 49.35    \\
                                 & VarRenamerNaive                 & 24.99    & 26.31   & 26.22    & 42.53    & 41.20   & 41.09    & 49.28 & 49.05 & 48.14    \\
                                 & VarRenamerRN                    & 14.75    & 15.41   & 15.78    & 31.65    & 31.48   & 30.56    & 37.07 & 37.07 & 36.03    \\
                                 & WorstCase                       & 1.46     & 8.59    & 10.67    & 2.43     & 22.86   & 21.92    & 2.58 & 27.66 & 27.70    \\ \hline
    \multirow{4}{*}{Format}      & Doc2Comments                    & 25.48    & 27.66   & 28.75    & 45.36    & 44.29   & 42.67    & 50.28 & 51.27 & 51.30    \\
                                 & NewLineInsertion \cellcolor{gray!30}                              & \cellcolor{gray!30} 20.44    & \cellcolor{gray!30} 22.32   & \cellcolor{gray!30} 25.54    & \cellcolor{gray!30} 35.52    & \cellcolor{gray!30} 37.86   & \cellcolor{gray!30} 37.91    & \cellcolor{gray!30} 39.74    & \cellcolor{gray!30} 46.19   & \cellcolor{gray!30} 48.58    \\
                                 & SplitLine                           & 27.07    & 28.42   & 30.04    & 44.60    & 45.18   & 43.44    & 52.07 & 52.16 & 51.83    \\
                                 & WorstCase                       & 15.96    & 18.58   & 21.58    & 32.00    & 34.50   & 34.36    & 35.80 & 42.18 & 44.87    \\ \bottomrule

    \end{tabular}

}
    \caption{Robustness evaluation for each category of perturbations on combined HumanEval and MBPP datasets. We highlight in gray the top four perturbation types that we have achieved the most improvements over the baseline $\lossclm$.}
    \label{tab:overall_granular_results}
\end{table*}

%% file: tables/ablation_full.tex
\begin{table*}
\setlength{\tabcolsep}{2.5pt}
\centering
\resizebox{\linewidth}{!}{
\begin{tabular}{l rr rr rr rr rrr}
\toprule
\multirow{2}{*}{Methods} & \multicolumn{2}{c}{Docstring} & 
\multicolumn{2}{c}{Function} & \multicolumn{2}{c}{Syntax} & \multicolumn{2}{c}{Format} & \multicolumn{3}{c}{Overall Average} \\ 
\cmidrule(lr){2-3} \cmidrule(lr){4-5} \cmidrule(lr){6-7} \cmidrule(lr){8-9}  \cmidrule(lr){10-12} 
& \npsk & \rpsk & \npsk & \rpsk & \npsk & \rpsk & \npsk & \rpsk & \npsk & \rpsk & \rdrop  \\
\midrule
$[0]\lossclm$ & 18.10 & 6.19 & 18.10 & 6.47 & 29.24 & 1.46 & 29.24 & 15.96 & 23.67 & 7.52 & 68.23\\
% $\lossdataaug$ & 18.28 & 7.93 & 18.28 & 7.41 & 30.56 & 9.35 & 30.56 & 19.12 & 24.42 & 10.95 & 44.84 \\
$[1]\lossdataaug$ & 18.10 & 7.45 & 18.10 & 7.94 & 30.11 & 8.59 & 30.11 & 18.58 & 24.10 & 10.64 & 55.85 \\
$[2]\lossmdataaug$ & 17.57 & 8.12 & 17.57 & 7.91 & 29.96 & 9.31 & 29.96 & 19.07 & 23.77 & 11.10 & 53.30\\
$[3]\lossmdataaug (p=0.2)$ & 17.91 & 7.36 & 17.91 & 7.84 & 31.30 & 8.88 & 31.30 & 17.86 & 24.60 & 10.48 & 57.40\\
$[4]\lossmbatchaug$ & 18.24 & 7.17 & 18.24 & 8.21 & 31.56 & 10.02 & 31.56 & 18.65 & 24.90 & 11.01 & 55.78\\
\rowcolor{gray!30}$[5]\lossmbatchaug + \lossalp$ & 18.03 & 7.38 & 18.03 & 8.19 & 31.30 & 11.92 & 31.30 & 18.44 & 24.67 & 11.48 & 53.47\\
$[6]\lossclm + \losscontraseq$ & 17.52 & 7.17 & 17.52 & 7.56 & 30.07 & 1.37 & 30.07 & 15.06 & 23.80 & 7.79 & 62.27 \\
$[7]\lossmbatchaug + \lossalp + \losscontratoken$ & 17.33 & 7.12 & 17.33 & 8.03 & 29.72 & 11.46 & 29.72 & 18.10 & 23.52 & 11.18 & 52.47 \\
$[8]\lossmbatchaug + \lossalp + \losscontraseq$ & 17.98 & 7.38 & 17.98 & 8.15 & 31.30 & 11.81 & 31.30 & 18.70 & 24.64 & 11.51 & 53.29 \\
$[9]\lossmbatchaug + \lossalp(\emph{On}+\emph{Co}) + \losscontratoken + \losscontraseq$ & 17.36 & 7.35 & 17.36 & 7.91 & 29.42 & 11.12 & 29.42 & 18.09 & 23.39 & 11.12 & 52.46\\
$[10]\lossmbatchaug + \lossalp(\emph{On}+\emph{Al}) + \losscontratoken + \losscontraseq$ & 17.31 & 7.33 & 17.31 & 7.80 & 29.77 & 11.48 & 29.77 & 18.12 & 23.54 & 11.18 & 52.51 \\
$[11]\lossmbatchaug + \lossalp(\emph{Bo}+\emph{Co}) + \losscontratoken + \losscontraseq$ & 16.87 & 7.19 & 16.87 & 7.72 & 29.40 & 11.07 & 29.40 & 17.82 & 23.14 & 10.95 & 52.68 \\
$[12]\lossmbatchaug + \lossalp(\emph{Bo}+\emph{Al}) + \losscontratoken + \losscontraseq$ & 16.70 & 7.15 & 16.70 & 7.77 & 29.54 & 11.41 & 29.54 & 17.62 & 23.12 & 10.99 & 52.47 \\
\bottomrule
\end{tabular}
}
   \caption{Ablation results of CodeGen-350M focusing on \emph{context-free perturbations}, i.e., we apply $\lossdataaug$ loss to the context-sensitive perturbations except for the baseline $\lossclm$.}
    \label{tab:ablation_results_free_full}
\end{table*}

\begin{table*}
\setlength{\tabcolsep}{2.5pt}
\centering
\resizebox{\linewidth}{!}{
\begin{tabular}{l rr rr rr rr rrr}
\toprule
\multirow{2}{*}{Methods} & \multicolumn{2}{c}{Docstring} & 
\multicolumn{2}{c}{Function} & \multicolumn{2}{c}{Syntax} & \multicolumn{2}{c}{Format} & \multicolumn{3}{c}{Overall Average} \\ 
\cmidrule(lr){2-3} \cmidrule(lr){4-5} \cmidrule(lr){6-7} \cmidrule(lr){8-9}  \cmidrule(lr){10-12} 
& \npsk & \rpsk & \npsk & \rpsk & \npsk & \rpsk & \npsk & \rpsk & \npsk & \rpsk & \rperct  \\
\midrule
$[0]\lossclm$ & 18.10 & 6.19 & 18.10 & 6.47 & 29.24 & 1.46 & 29.24 & 15.96 & 23.67 & 7.52 & 31.77\\
$[1]\lossmbatchaug$ & 18.68 & 8.19 & 18.68 & 8.56 & 30.98 & 7.72 & 30.98 & 18.54 & 24.83 & 10.75 & 43.29\\
\rowcolor{gray!30}$[2]\lossmbatchaug + \lossalpdrop$ & 18.80 & 8.44 & 18.80 & 8.37 & 30.84 & 8.00 & 30.84 & 18.98 & 24.82 & 10.94 & 44.08 \\
$[3]\lossmbatchaug + \losscontraname$ & 18.65 & 8.14 & 18.65 & 8.49 & 30.54 & 7.94 & 30.54 & 18.91 & 24.60 & 10.87 & 44.19\\
$[4]\lossmbatchaug + \lossalpdrop + \losscontraname$ & 18.63 & 8.12 & 18.63 & 8.26 & 30.69 & 7.82 & 30.69 & 18.98 & 24.66 & 10.80 & 43.80 \\
$[5]\lossmbatchaug + \lossalpdrop + \losscontraname + \losscontratoken$ & 18.31 & 7.79 & 18.31 & 7.80 & 30.53 & 8.17 & 30.53 & 17.93 & 24.42 & 10.42 & 42.67 \\
$[6]\lossmbatchaug + \lossalpdrop (\emph{All})$ & 18.07 & 8.66 & 18.07 & 8.12 & 30.33 & 7.56 & 30.33 & 18.88 & 24.20 & 10.80 & 44.63 \\
$[7]\lossmbatchaug + \lossalpdrop (\emph{AllS})$ & 18.37 & 8.12 & 18.37 & 8.73 & 30.51 & 7.49 & 30.51 & 18.08 & 24.44 & 10.61 & 43.42 \\
\bottomrule
\end{tabular}
}
\caption{Ablation results of CodeGen-350M focusing on \emph{context-sensitive} perturbations, i.e., we apply $\lossdataaug$ loss to the context-free perturbations except for the baseline $\lossclm$.}
    \label{tab:ablation_results_sensitive_full}
\end{table*}